\documentclass[sn-mathphys]{sn-jnl}

\usepackage{graphicx}
\usepackage{subcaption}

\makeatletter
\renewcommand{\ALG@name}{Pseudo-code}
\makeatother

\jyear{2021}%

\theoremstyle{thmstyleone}%
\theoremstyle{thmstyletwo}%

\theoremstyle{thmstylethree}%

\begin{document}

\title[Congestion Management in Interconnects using Adaptive Routing]{Congestion Management in High-Performance Interconnection Networks Using Adaptive Routing Notifications}


\author[1]{\fnm{Jose} \sur{Rocher-Gonzalez}}\email{jose.rocher@uclm.es}

\author[1]{\fnm{Jesus} \sur{Escudero-Sahuquillo}}\email{jesus.escudero@uclm.es}

\author[1]{\fnm{Pedro J.} \sur{Garcia}}\email{pedrojavier.garcia@uclm.es}

\author[1]{\fnm{Francisco J.} \sur{Quiles}}\email{ francisco.quiles@uclm.es}

\affil[1]{\orgdiv{Department of Computing Systems}, \orgname{Universidad de Castilla-La Mancha}, \orgaddress{\street{Campus Universitario}, \city{Albacete}, \postcode{02071 }, \state{Castilla-La Mancha}, \country{Spain}}}

\abstract{The interconnection network is a crucial subsystem in High-Performance Computing clusters and Data-centers, guaranteeing high bandwidth and low latency to the applications' communication operations. Unfortunately, congestion situations may spoil network performance unless the network design applies specific countermeasures. Adaptive routing algorithms are a traditional approach to dealing with congestion since they provide traffic flows with alternative routes that bypass congested areas. However, adaptive routing decisions at switches are typically based on local information without a global network traffic perspective, leading to congestion spreading throughout the network beyond the original congested areas. In this paper, we propose a new efficient congestion management strategy that leverages adaptive routing notifications currently available in some interconnect technologies and efficiently isolates the congesting flows in reserved spaces at switch buffers. The experiment results based on simulations of realistic traffic scenarios show that our proposal removes the congestion impact.}




\maketitle

\section{Introduction}\label{sec1}
The use of High-Performance Computing (HPC) systems and Data-centers has grown to face today's challenges of applications and services used in fields such as Genomics, Climate Forecast, Robotics, Artificial and Business Intelligence, Cloud Computing, Big-Data, etc.
These services and applications are decomposed in tens of thousands processes, whose execution demands a fast communication among the system processing and storage nodes. Hence, these nodes are interconnected through a high-performance interconnection network, that must offer mainly high bandwidth and low latency, otherwise becoming the system bottleneck. Therefore, the design of the interconnection network is crucial to meet the system requirements.

In that regard, some key features of the interconnection network design are cost-effective network topologies, efficient routing algorithms, and techniques that guarantee the performance under unexpected traffic conditions or faults.
Regarding topologies, the Fat Tree \cite{FatTree} is widely used both in HPC systems (like in many supercomputers ranked in the TOP500 list) and in Data-centers\footnote{Note that the term CLOS is currently employed, mainly in the Data-centers context, to refer to topologies similar (if not identical) to Fat Trees. Although the original CLOS network was a three-stage non-blocking unidirectional topology, the fact that a Fat Tree can be built by recursively applying the CLOS construction method (followed by folding) led to using the terms CLOS and Fat Tree interchangeably.}.
Fat-Tree-like topologies, among other benefits, offer a high number of alternative paths between two endnodes (i.e., path diversity), leading to inherent fault tolerance capabilities.

The fat-tree networks properties are leveraged by different routing algorithms, either deterministic \cite{Gomez07ipdps,Zahavi2010}, oblivious \cite{ObliviousFT} or adaptive \cite{DBLP:journals/jsac/ZahaviKK14,DBLP:conf/hoti/GeoffrayH08,DBLP:conf/sc/KimDA06, adaptiveRCA,adaptiveDBAR}. 
While deterministic routing algorithms always select the same path to communicate two endnodes, oblivious routing chooses one of the available paths randomly \cite{ecmp, presto}, without any information of the network traffic status.
By contrast, adaptive routing selects among the available routes the most proper one based on the switch buffer occupancy.
This means that a route will be selected if the buffers along that path contain the lowest number of packets among all the available routes.
Based on this idea, there are techniques based on monitoring local switch buffer occupancy to select the less loaded paths (local information) \cite{drill, power2}.
Nowadays, adaptive routing schemes are widely used in commercial interconnection networks, such as InfiniBand or Slingshot.


Furthermore, adaptive routing proposals using regional or global information for path selection have been proposed \cite{DBLP:journals/jsac/ZahaviKK14, adaptiveRCA, adaptiveDBAR}.
Minimal-path adaptive routing selects one of the available minimal paths (i.e., the cost in terms of hops to reach the destination is the smallest one), based on some local or regional information about the clogged paths in the network.
There are many network topologies, such as fat-trees, dragonflies or Tori, which exploit minimal-path routing. 
By contrast, non-minimal adaptive routing selects between one or more minimal paths and one or more non-minimal paths. 
Note that the non-minimal path selection makes sense to increase bandwidth between endpoints or to avoid hot-spots in the minimal path. 
For instance, Slim-fly \cite{Besta14sc}, Jellyfish \cite{Jellyfish} and Dragonflies \cite{Kim08isca} are modern low-diameter network topologies that use non-minimal routing either adaptive, such as Universal Globally-Adaptive Load balanced (UGAL) \cite{Kim08isca}, or oblivious, such as Valiant Load Balancing (VLB) \cite{Valiant82siamcomp}.
While UGAL uses the information in the source router of a network path, there are other non-minimal adaptive routing solutions that base their decision on the information that is not directly available at the source router.
For instance, progressive adaptive routing (PAR), piggyback routing, and reservation routing \cite{Jiang09, Newaz21} are based on this approach.
In general, it is not our intention in this paper to analyze all the adaptive routing proposals, since there has been an extensive research in this field.
Further details on adaptive routing mechanisms can be found in several surveys, such as the Besta et al. work \cite{multiPathTable}.

Although there are many efficient topologies and routing algorithms that have been thoroughly designed for large HPC and Data-centers, the network congestion generated by intensive communication operations of applications and services may severely degrade the network performance \cite{Congdon21Nendica}.
Network congestion occurs when several traffic flows simultaneously and persistently request access to the same network resource (e.g. endnode, buffer or device port).
This may happen, for instance, due to ``many-to-one'' communication operations, generated by several endnodes sending traffic to a single one (\emph{incast} congestion), but also due to traffic flows contending at the inner part of the network (\emph{in-network} congestion).
In lossless networks, where packet dropping is not allowed, congestion may be eventually propagated backwards throughout the network by the effect of the flow-control back-pressure. This creates structures of blocked packets, usually referred to as congestion trees \cite{Garcia05hipeac}, whose \emph{root} is located in the switch port where congestion originates (i.e, where traffic flows collide). Analogously, the tree \emph{branches} have a shape determined by the paths where congestion propagates towards the sources, or, in the opposite direction, by the paths followed by the flows contributing to create the root (congesting flows). Finally, the tree \emph{leaves} are the ends of the branches.  

When congestion appears, it usually produces two negative effects: the \emph{Head-of-Line (HoL) blocking} \cite{DBLP:journals/tcom/KarolHM87} and the \emph{buffer hogging} \cite{bufferhogging}.
HoL blocking is likely to appear at the switch where the root of a congestion tree is located, when a packet at the head of an input queue is stuck while trying to access the root, then blocking the packets stored behind in that queue, even if they request other output ports in that switch.
This situation is called \emph{low-order HoL blocking} \cite{DBLP:journals/tcom/KarolHM87}.
In addition, \emph{high-order HoL blocking} \cite{Jurczyk96phenomenonof} 
may occur when congestion propagates to other switches different to that where it originates, then the congestion-tree branches blocking non-congesting packets.
On the other hand, the \emph{buffer hogging} \cite{bufferhogging} appears when congesting flows hog most of the buffer space at a given switch port, so throttling the bandwidth rate available for non-congesting flows. 
Both HoL blocking and buffer hogging may dramatically degrade the network performance, hence implementing some strategy to deal with congestion or its effects is almost mandatory in high-performance interconnects.

One of the traditional approaches to deal with congestion is adaptive routing, so that the traffic flows are routed through alternative routes that bypass congested areas.
However, as we showed in previous studies \cite{RocherJPDC21,rocherCCgrid}, this approach may end up spreading congestion through network paths, thus worsening the congestion situation as more network routes are clogged. Congestion spreading increases the negative impact of congestion trees, since more branches are added to these tress when congesting flow are sent through alternative routes.
Moreover, congestion spreading also amplifies the negative effects of congestion (i.e., HoL blocking and buffer hogging), since their appearance is increased when congestion spreads.
Congestion spreading also spoils the performance of techniques that try to reduce HoL blocking based on statically separating flows into different queues (or virtual channels, VCs) \cite{Nachiondo10pds,Guay11ipdps,Escudero14jpdc}, as the spread congesting flows are likely to be present in more queues (see section \ref{sec:background:congestion_management}). 

To reduce the congestion spreading and its derived problems, we proposed  \emph{Adapted-Flow Isolation} (AFI) \cite{rocherCCgrid}, a technique that isolates the re-routed (or ``adapted'') packets in a virtual channel different to that used by non-adapted packets. In practice, AFI separates non-congesting flows from congesting ones that are spread throughout the network due to adaptive routing, so saving the former from suffering HoL-blocking and buffer hogging due to the latter (see section \ref{sec:background:AFI}).
However, AFI assumes that congesting flows are re-routed (so eventually isolated) based on local switch information, which actually is not enough to avoid high-order HoL blocking.
The main shortcoming of AFI is that it spreads the congesting flows among all the alternative paths when a switch detects congestion. Although the congesting flows are isolated on a special VCs, the AFI performance could be improved if we reduce congestion spreading (though isolated in special VCs) to happen in a smaller number of links.

Indeed, AFI could be improved if switches share congestion information, so that they are able to identify and isolate the congesting flows based on a more accurate knowledge of the network status and congestion state.
For this reason, it is desirable that switches identify the congesting flows and send congestion information notifications to neighboring switches, so that they can made better adaptive routing decisions.
Note that the last generation of InfiniBand-based switches support congestion notifications between switches.
Specifically, these switches generate adaptive routing notifications (ARNs) \cite{MellanoxQuantum, MellanoxConfigureAR} upon congestion or link failure detection\footnote{Note that a link failure may also generate a congestion situation, as the network bisection bandwidth decreases due to this failure.}.
ARNs are sent to neighboring switches, which may consume or forward them upstream to other switches.

In this paper, we propose a new congestion management technique that reduces the congestion spreading (even more than the AFI proposal) produced by adaptive routing algorithms, so that the congestion impact in the network is reduced significantly as congesting flows are identified more accurately.
Our new proposal leverages the ARNs support, available in some commercial switches, so that adaptive routing decisions are made based on the information exchanged by neighboring switches.
For the sake simplicity, we have evaluated our approach in networks using fat-tree topologies, since they are commonly used in real Supercomputers and Data-center.
However, our proposal could be applied to other topologies where congestion spreading happens.
On the other hand, in this paper we provide some implementation details for InfiniBand-based hardware, since there is information made publicly available of their switches architecture that we have used to adapt our simulation model.
Notwithstanding, our proposal could be applied to other technologies as long as they provide support for virtual channels, adaptive routing, and congestion notifications between switches. In summary, the main contributions of this paper are the following:

\begin{enumerate}

\item We propose a new technique that is aware of the congestion wherever it originates,  and  leverages  the  congestion notifications among switches to identify congesting flows.

\item We define a congestion detection mechanism assuming a switch architecture that combines the use of virtual channels (VCs) and virtual output queues (VOQs).

\item We leverage the adaptive routing notifications (ARNs) mechanism, available in current InfiniBand networks, so that, when congestion is detected at a given switch port, this switch propagates the congestion information by means of ARN packets, so that neighboring switches adapt the congesting flows to sidestep the congestion root.

\item We use the AFI approach to isolate the adapted flows in special buffer resources at switches, so that they are routed, thereafter, using deterministic routing. In this way, congestion spreading is reduced when adaptive routing is used.

\item We provide implementation details of our approach in InfiniBand-based networks, using the ARNs mechanism. 

\item We evaluate the behavior of our proposal with different queuing schemes, in order to study the impact of providing an extra Head-of-line (HoL) blocking reduction under different congestion scenarios.

\end{enumerate}

The rest of this paper is organized as follows.
Section \ref{sec:background} overviews the essential background.
Section \ref{sec:implementation} describes our proposal, which is evaluated based on simulations in Section \ref{sec:evaluation}.
Finally, in Section \ref{sec:conclusions} some conclusions are drawn.


\section{Background}
 \label{sec:background}
 
\subsection{Fat-Tree Networks and Routing}
 \label{sec:background:fat-trees}
 
The Fat-Tree family includes several multi-stage topologies that differ mainly in the number and arrangement of links between stages.
This arrangement (or connection pattern) determines the bisection bandwidth of the topology, being constant, slimmed or fat, such that we can distinguish between Real Life Fat-Trees (RLFTs) \cite{DBLP:conf/ipps/Zahavi11}, slimmed Fat-Trees \cite{slimmedFT}, ``Fat'' Fat-Trees, etc.

RLFTs are a sub-class of Parallel Ports Generalized Fat-Trees (PGFTs), which have been widely adopted by the industry, due to their reduced cost per switch-port ratio.
In the RLFTs, all the switches have the same ports count $P$, and every switch connects $K$ ports ($K=P/2$) to other switches in the next stage, and $K$ ports either to other switches or to endnodes in the previous stage.
An important property of RLFTs is that if $K$ remains constant regardless the stage, the network bisection bandwidth is constant.
Furthermore, the number of end nodes ($N$) interconnected by a $n$-stage RLFT with $K=P/2$ is given by $N=2(K^n)$, and the number of switches ($S$) can be calculated as $S=(N\cdot(2n-1))/2K$.

RFLTs offer multiple routes and shortest-paths diversity, that are leveraged by several routing algorithms, either deterministic, oblivious or adaptive.
In general, routing in Fat-Trees is divided in three phases: upward, turnaround and downward.
First, in the upward phase, one among the $K$ output ports of a switch connecting to the upper stage can be selected to route a packet.
This may be repeated at several switches, until the packet reaches the switch where it turns around to be routed through an output port connecting to the lower stage. Then, in the downward phase, the packet is routed through a specific single path to reach its destination, regardless the routing algorithm.
Note that the set of routes offered by RLFTs is limited by $K$ and by the number of stages $n$.

There are several deterministic routing algorithms proposed for RLFTs, such as $D$-mod-$K$ \cite{Zahavi10cpe,Gomez07ipdps}, which smartly balances the traffic flows through the available minimal paths, such that all the links between any given two stages are used by the same number of minimal paths. 
$D$-mod-$K$ achieves the same performance as that offered by oblivious and adaptive routing, under certain traffic conditions (e.g., when traffic flows destinations follow a uniform distribution).
However, $D$-mod-$K$, like any deterministic routing algorithm, is not able to react to congestion situations and their negative effects (e.g., HoL blocking) once they appear.
By contrast, adaptive routing algorithms proposed for Fat-trees, are able to select alternative routes
when congestion appears. 

HPC interconnect technologies such as InfiniBand, Slingshot, OmniPath, and BXI offer different approaches to adaptive routing. Intel Omnipath adapts the routing when a switch detects that a route has failed (due to congestion or failure), then it selects the route from the list of alternative routes and updates its forwarding table \cite{OmnipathFM}. The switches use a threshold parameter to determine when a route is busy based on the buffer occupancy. Alternative route selection is tunable, and it offers "Random" (choose the alternate randomly), "Greedy" (choose the least busy alternative) or "GreedyRandom" (if there are multiple alternates that are not busy, randomly choose from them).
BXIv2 adaptive routing \cite{BXIV2} feature is targeted towards improving communication performance but not fault management. In BXI, a message is a candidate for adaptive routing only if the in-order guarantee is not required and if its size is above a customizable threshold.  If the message is adaptable, then the switch picks several output ports randomly. These are the candidates, and the algorithm compares against the deterministic output port which one of these has the lowest congestion, based on the buffer occupancy, and selects that port to route the traffic.
HPE Slingshot routes packets dynamically according to load \cite{slingshot}.
The congestion is estimated by considering the total depth of the request queues of each output port. This congestion information is distributed on the chip by using a ring to all the forwarding blocks of each input port.
Before sending a packet, the source switch estimates the load of up to four minimal and non-minimal paths (not applicable in fat-tree) and sends the packet on the best path, which is selected by considering both the paths’ congestion and length. 

Infiniband adaptive Routing \cite{MellanoxConfigureAR}  enables the switch to select the output port based on the buffer occupancy. It assumes no constraints on output port selection (free AR).  This enhancement can be activated together with Self-Healing Networking which enables the switch to select the alternative output port without Subnet Manager intervention. Fast Link Fault Notification (FLFN) and Adaptive Routing Notification (ARN) enable the switch to report to neighbour switches the status of the switch when it is not able to select an alternative path. 

Table \ref{tab:hw} summarizes the differences between the different network technologies.

\begin{table}[h]
    \centering
    \begin{tabular}{ | r| c | c | c}
    \hline
         Technology & Trigger &  Trigger Reaction\\
         \hline
         Intel Omnipath & local status & Random or greedy or greedyrandom routing \\
         BXI v2 & local Status & Greedy routing, random subset of valid output ports.\\
         HPE Slingshot & global status & Greedy routing (congestion and depth)  \\
        Infiniband & ARN & Greedy routing avoiding congested paths  \\
        \hline
    \end{tabular}
    \caption{Summary of adaptive routing techniques in HPC network technologies.}
    \label{tab:hw}
\end{table}

 \subsection{Adaptive Routing in InfiniBand}
 \label{sec:background:arn}

The InfiniBand Architecture (IBA) specification \cite{IBAspec2015} 
defines the multi-path routing, when the topology supports it, by assigning multiple local identifiers (LIDs) to the same endnode.
Indeed, routing tables at switches are populated using the multiple LIDs of a specific endnode, so that different paths in the topology could be selected to reach that endnode.
However, the policy to select the paths is actually oblivious. Thus the switch is not aware of the network traffic status, and the link or buffer occupancy.

Fortunately, there are alternative adaptive routing implementations available in commercial IBA devices, such as those manufactured in the last few years by Mellanox/NVIDIA.
Specifically, a patent filed by Mellanox in 2015 defined \emph{Adaptive Routing Notifications} (ARNs) \cite{patenteARN}, a mechanism that uses the local switch information to detect congestion (or network faults), and then generates special notification packets (the ARNs) to notify the neighboring switches about this congestion scenario\footnote{Thanks to the ARNs mechanism, the switches also monitor the switch link status to detect link failures. Hereafter, we will only focus on the congestion monitoring functionality.}.
Based on the received ARNs, neighboring switches may select alternative routes whenever is possible, thus consuming these ARNs.

\begin{figure}[!h]
\centering
\includegraphics[width=0.75\columnwidth]{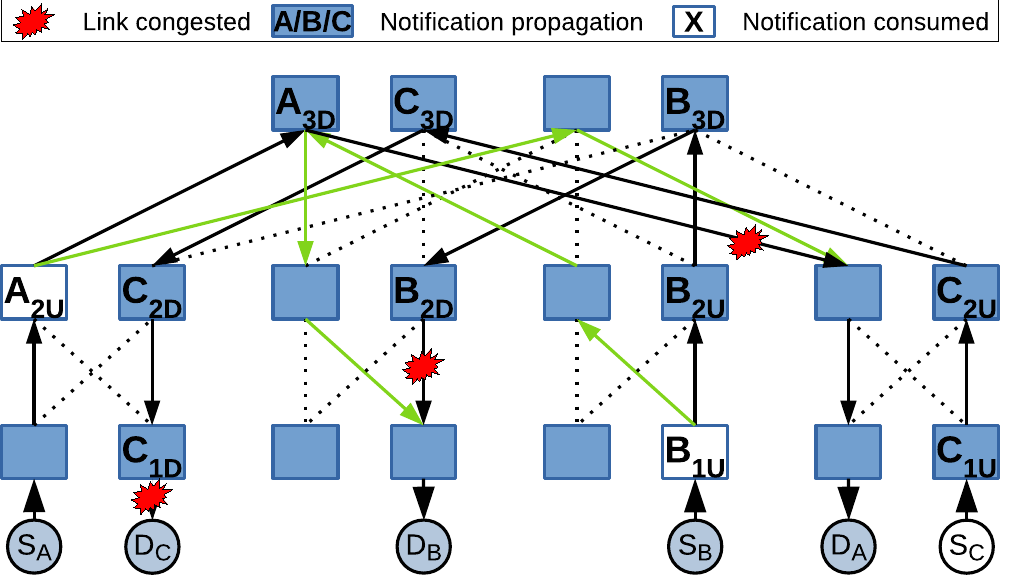}
\caption{Example of a $3$-stage Fat-Tree using ARNs.}
\label{fig_rules}
\end{figure}

Note that whether the switch can consume or not the ARN depends mainly on the network topology.
For instance, Fat-tree networks allow routing algorithms to select multiple paths in the upward phase, but not in the downward phase, as there is only one possible path to reach an endnode after the turnaround phase (see Section \ref{sec:background:fat-trees}).
Hence, if there is a congestion situation in a downward path, packets following that path must be adapted at some point of the upward phase where it is possible to select an alternative route that avoids crossing the congestion root.
Therefore, the switch consuming the ARN is determined by its stage in the Fat-Tree topology, which must be one stage lower than the stage of the switch generating the ARN.
\figurename~\ref{fig_rules} shows an example containing three possible scenarios of ARN generation and consumption in a 3-stage Fat-tree.

We show three traffic flows $A, B, C$, between source and destination endnodes, denoted by $S_f$ for the endnode generating flow $f$, and $D_f$ for the endnode receiving that flow ($f$ being $A$, $B$ or $C$).
Flow $A$ communicates endnodes $S_{A}$ and $D_{A}$, and suffers a congestion situation in a downward path, between the third and second stages.
The ARN is generated in the switch of the third stage (denoted by $A_{3D}$ and colored in blue) and needs to reach the switch in the second stage ($A_{2U}$ in white) to be consumed.
In switch $A_{2U}$, packets of flow $A$ can be routed in the upward phase selecting an alternative path, among those colored in green, to avoid the congested switch $A_{3D}$.
Flow $B$ communicates $S_{B}$ and $D_{B}$, and it suffers congestion between the second and first stages in the downward phase (switch $B_{2D}$), hence the ARN has to reach the first stage (switch $B_{1U}$) to be consumed.
In switch $B_{1U}$, packets of flow $B$ will be routed selecting the alternative upward path colored in green.
For flow $C$, between $S_{C}$ and $D_{C}$, the congestion originates between switch $C_{1D}$ in the first stage and endnode $D_C$ (i.e., an incast situation), so the ARNs will not be consumed until endnode $S_C$.
From $S_C$ there is no alternative route to reach $D_C$, which avoids the congested switch $C_{1D}$, so packets from flow $C$ will be unavoidably those contributing to create congestion.
Indeed, the ARN mechanism prevents switches from adapting traffic flows creating \emph{incast} congestion, since the ARNs generated in these situations will be consumed by the endnodes.
By contrast, the ARNs mechanism does not prevent \emph{in-network} congestion.

\subsection{Static Queuing Schemes}
 \label{sec:background:congestion_management}

Techniques based on Static Queuing Schemes (SQSs) reduce the HoL blocking derived from congestion situations by mapping different packet flows to different queues, Virtual Channels (VCs), or Virtual Lanes (VLs) in the InfiniBand nomenclature. These mapping is performed ``a priori'', i.e., it does not depend on the network traffic state.
In this way, the sharing of buffer space among flows is reduced as much as possible with the number of available queues per port. 

There are several examples of SQSs specially tailored to Fat-Trees using deterministic routing (e.g., $D$-mod-$K$ \cite{Gomez07ipdps, Zahavi2010}).
For instance, vFtree \cite{Guay11ipdps} takes into account the source and destination leaf switch of each packet, and shuffles the packets addressed to consecutive leaf switches among the available queues in a switch buffer. 
Another scheme is Flow2SL \cite{Escudero14jpdc}, which defines as many groups of consecutive destinations as the number of available queues,
and maps packets to the different queues or VLs depending on their source and destination group.
On the other hand, there are other proposals agnostic to the network topology and routing, such as \emph{Destination-Based Buffer Management} (DBBM) \cite{Nachiondo10pds}.
This scheme uses a module function to map packets with the same destination to the same queue in any switch port. 
Note that, as the number of available queues (or VCs or VLs) is always lower than the number of destinations, all the schemes mentioned above allow that subsets of destinations share queues at some network ports.

While the approach followed by the SQSs is able to reduce significantly the HoL blocking when congestion appears, we reported in several studies \cite{rocherCCgrid,RocherJPDC21} that these proposals have several limitations when adaptive routing is used, due to the congesting flows spreading throughout the network, which leads them to share more queues with non-congesting flows.
A possible solution for this problem is tweaking adaptive routing operation so that the adaptivity degree is partially restricted
and ultimately congestion spreading is reduced. However, this approach neither identifies exactly the congesting flows, nor considers the network traffic conditions to adapt specific traffic flows.

\subsection{Adapted-Flow Isolation (AFI)}
 \label{sec:background:AFI}

The \emph{Adapted-Flow Isolation} (AFI) technique \cite{rocherCCgrid} aims to solve the problem of  adapted congesting flows sharing queues with non-congesting ones in more switch buffers than in deterministic routing scenarios. Specifically, AFI assumes that congestion can be detected, and that congestion detection triggers the selection of alternative routes. In these cases, AFI marks as adapted the re-routed packets that are, thereafter, isolated in a special virtual channel, called \emph{Adapted-Flow Channel} (AFC). A single AFC per switch port is required. Once stored in the AFC, these packets will be always routed following a deterministic route (i.e., they will not be adapted again), thus reducing the congestion spreading in the alternative paths. As the non-adapted packets are stored in standard VCs, different to the AFC, adapted flows never produce HoL-blocking to non-adapted ones. Note that the standard VCs can be used to implement any static queuing scheme, whose performance will not be spoiled by congesting flows if they are adapted and so isolated in the AFC. However, note that AFI isolates in the AFC all the re-routed packets, but re-routing decisions are made based on some congestion detection mechanism. Therefore, AFI efficiency depends ultimately on the accuracy of the congestion detection mechanism. In other words, AFI would achieve maximum efficiency if all and only the actual congesting flows were re-routed, but this requires a precise identification of such flows. This is almost impossible to achieve if congestion detection is based only on local information, especially in high-order HoL-blocking situations. Hence, in this paper we propose a solution for this problem based on the use of ARNs, as explained in the next section.

\section{Adapted-Flow Isolation using ARNs}
 \label{sec:implementation} 

In this section, we present ARN$+$AFI, a new technique for congestion management in high-performance interconnection networks using adaptive routing.
In the following sections, we describe the basic operation details, the assumed switch architecture, the procedure for congestion detection and isolation, and the implementation details.

\subsection{Basic Details}
 \label{sec:implementation:basics}

The main objective of ARN$+$AFI is to reduce as much as possible the congestion spreading and their negative effects (i.e., HoL blocking and buffer hogging), when adaptive routing is used.
To achieve this goal, ARN$+$AFI leverages the Adaptive Routing Notification (ARN) mechanism (see Section \ref{sec:background:arn}) to propagate the congestion trees information to neighboring switches, and the Adapted-Flow Isolation (AFI) technique \cite{rocherCCgrid} (see Section \ref{sec:background:AFI}) to identify and isolate the congestion flows being adapted much more accurately. Specifically, ARN$+$AFI uses a congestion detector (see Section \ref{sec:implementation:detector}) at every switch to detect congestion roots.

When a packet at some input port requests the congested output port, an ARN is sent from that input port to upstream switches (see Section \ref{sec:implementation:arn}).
A switch will ``consume'' an ARN when it finds an alternative path in the topology to route packets addressed to the destination contained in the ARN (i.e., to sidestep the congestion root), otherwise this switch will propagate the ARN to upstream switches.
Note that ARN$+$AFI isolates the packets being routed through alternative routes in a special virtual channel at switch ports, called Adapted Flow Channel (AFC) as defined by AFI (see Section \ref{sec:background:AFI}), regardless of whether congestion is detected locally or via consumed ARNs from downstream switches.
Since a packet is stored in the AFC must be stored in the AFC of the subsequent switches along its route in the network to prevent packets routed through alternative routes interacting with other packets routed through default routes.
Indeed, ARN$+$AFI routes the packets in the AFC using deterministic routing, thus reducing the congestion spreading.
ARN$+$AFI can be also combined with static queuing schemes (SQS) in order to reduce even more the HoL blocking and buffer hogging.

 \subsection{Assumed Switch Architecture}
 \label{sec:implementation:architecture}

\begin{figure}[!h]
\centering
\includegraphics[width=0.7\columnwidth]{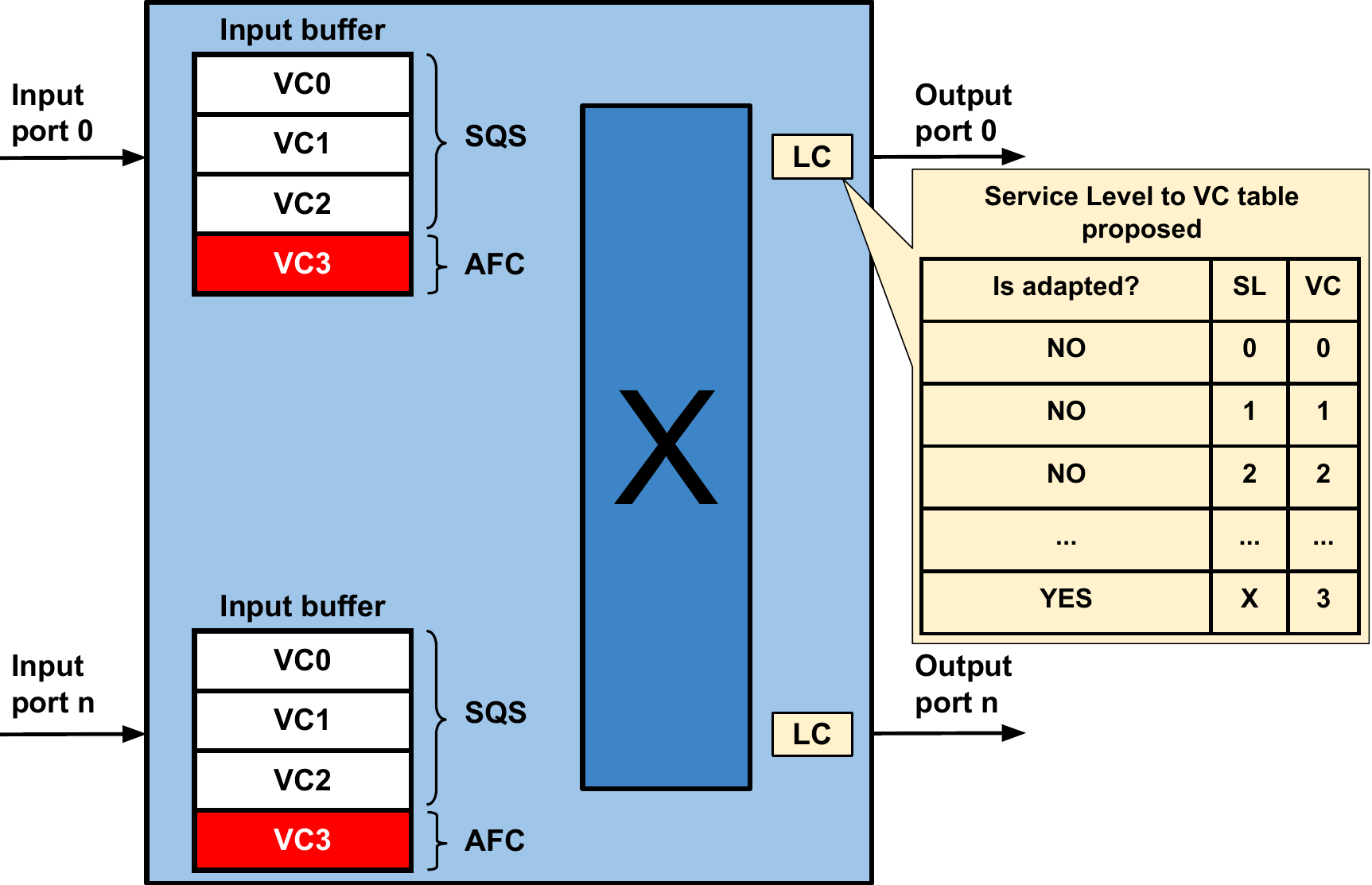}
\caption{Diagram of an $n$-port IQ switch. Buffers are divided in virtual channels (VCs). Flow-control is performed at VC level. The VOQs operation is shown in Figure \ref{fig_congestion}.}
\label{fig_sw_architecture}
\end{figure}

Hereafter, we assume Input Queued (IQ) switches, which are commonly used in high-performance interconnection networks.
IQ switches place the buffers at input ports, and they allow to combine the use of flow-controlled virtual channels (VCs) with a congestion detection procedure, as required by ARN$+$AFI.
Indeed, to perform an accurate congestion detection, we also assume virtual output queues (VOQs) at input buffers, operating together with the VCs in an orthogonal way.
VOQs measure the occupancy of output ports and the contribution of every input port to this occupancy, since we want to both detect congestion and identify the input ports (i.e., buffers) more clogged.
\figurename~\ref{fig_sw_architecture} shows the assumed switch architecture. 

We define at least two VCs at each switch input port: one to store non-adapted packets and another (i.e., the AFC) storing adapted ones.
Also, note that we could use static queuing schemes (SQSs) combined with our proposal.
In this case, the switch needs additional VCs, apart from the AFC, to implement the SQS (see Section \ref{sec:background:congestion_management}).
We assume the AFC is always the last one of the available VCs.
Moreover, two switches connected through a link perform credit-based flow control at VC-level.  
The link controller (LC) is aware of the number of credits available per VC in the next switch input port.
Packets are mapped to their corresponding VC based on the \emph{Service Level to VC table} at the LC (see \figurename~\ref{fig_sw_architecture}).
This table defines to which VC a packet will be mapped in the next switch, based on the traffic class (TC) or service level (SL) identifier, stored in the packet header.
If a packet is already ``adapted'' (i.e., the ``adapted'' bit in its header is set), then it will be mapped the AFC, regardless of its TC or SL identifier.

As mentioned above, we assume VOQs at input buffers to measure the output ports occupancy.
However, unlike the VCs, the VOQs at a given port are not flow controlled, so that the occupancy of a given VOQ may grow in congestion situations, then hogging the entire buffer space of that port.
As we assume at least two flow-controlled VCs, a growing VOQ will only hog the space of one out of the two VCs.
Moreover, most of the switch implementations reserve a minimal number of memory slots to each VC, so that the maximum size of a VOQ is bounded to the free memory slots not reserved for the VCs.

Regarding the routing algorithm, deterministic routing is used by default (e.g., $D$-mod-$K$ in Fat-trees).
Each switch output port uses a round-robin arbiter that gives the same priority to all the input ports, and, at each input port, it also gives the same preference to all the available VCs (i.e., the regular VCs and the AFC).

\subsection{Congestion Detector}
 \label{sec:implementation:detector}



ARN$+$AFI assumes a congestion detector at switches that monitors the buffers occupancy, detects congestion just when it originates in any network point, differentiates a congested switch port being either the root of a congestion tree or part of a growing congestion tree branch, and notices if a congestion tree root remains in a switch port for longer or it moves downstream after a short period of time.
First, the congestion detector monitors the switch buffers occupancy (i.e., the VOQs) at every input port.
When the occupancy of a VOQ exceeds the \emph{high congestion detection threshold} (HCDTh), then the output port to which that VOQ forwards packets is considered as a potential congestion root.
In this moment, we assume that the packet at the head of that VOQ is the responsible of this congestion situation.
Note that the size of a VOQ may hog several VCs, since VOQs are not flow controlled (see Section \ref{sec:implementation:architecture}).

The next step is to figure out whether the output port exceeding the HCDTh is the root of a congestion tree or part of a congestion tree branch being propagated to the current switch.
For this purpose, the congestion detector asks the link controller (LC) of that output port for  the number of available credits in the next switch VC where the packet responsible of congestion will be stored.
If this number is higher than the \emph{free credits threshold} (FCTh), then this output port is the root of a congestion tree, otherwise it is part of a congestion tree branch\footnote{We assume the same criterion as defined in Section A10.1.2 of the InfiniBand specification.}. Note that an output port being the root of a congestion tree receives packets faster than it is able to forward, and it has room in the next switch buffer to store congesting packets. However, if the congested output port is part of a congestion tree branch, the available credits in the next buffer will be near zero, as congestion being propagated makes the flow control mechanism to back-pressure upstream switches contributing to congestion.

When the HCDTh is exceeded and the number of free credits in the next switch port is higher than FCTh, the congestion detector has to check that the congestion tree root remains in that output port for longer, and it does not move downstream after a short period of time. For this purpose, we define a timer, called \emph{congestion root timer} (CRT), so that, when it expires and the HCDTh and FCTh are exceeded, the output port is finally considered as a congestion root.

\begin{figure}[h!]
\centering
\includegraphics[width=0.75\columnwidth]{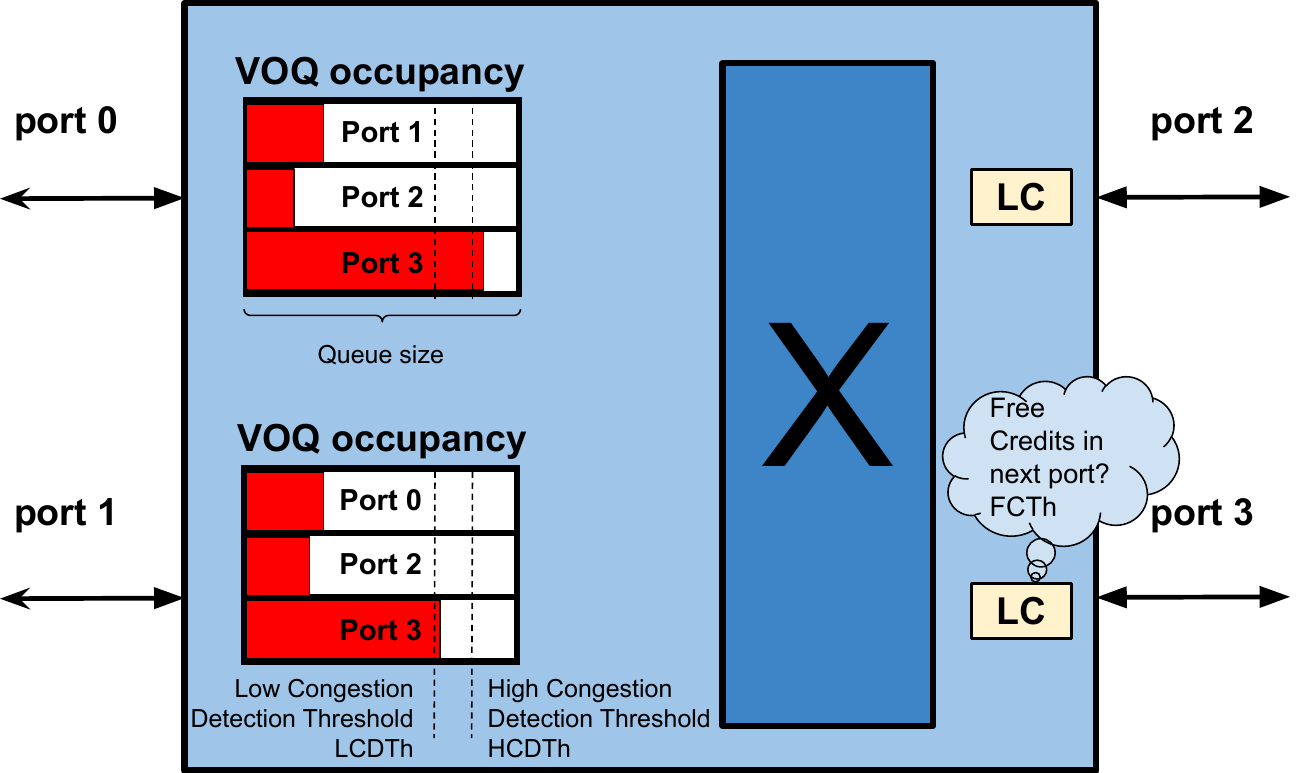}
\caption{Example of a congestion detection in a switch.}
\label{fig_congestion}
\end{figure}

\figurename~\ref{fig_congestion} shows an example of the operation of the congestion detector. For the sake of clarity we omit the operation of VCs in this example.
Note that the occupancy of VOQ \#3 at port \#0 exceeds the \emph{HCDTh} since packets stored at this VOQ contend with those stored in VOQ \#3 at port \#1 for the output port output port \#3.
Therefore congestion is detected at port 0, and we need to ask the LC at port \#3 (i.e., the port forwarding packets from port 0, VOQ \#3) if the available credits in the next switch exceed the \emph{FCTh} for the VC where the packet responsible of congestion will be stored.
If these two conditions are satisfied, the congestion detector will wait for the \emph{CRT} to expire. When the timer expires, if the two previous conditions are still true then the output port \#3 will be identified as the root of the congestion.
At this moment ARN$+$AFI will be generate an ARN based on the information of the packet being responsible of congestion.

When congestion disappears, the congestion detector disables the output port as the root of a congestion tree.
As different input ports VOQs may have exceeded the \emph{HCDTh}, the congestion root is disabled only when the occupancy of all those VOQs is below the \emph{low congestion detection threshold} (LCDTh).
As more than one VOQ from different input ports may exceed the \emph{HCDTh}, we need to keep this number and check that it reaches zero in order to disable that congestion root.
This approach is followed by InfiniBand networks \cite{GranZRSSL11}.

\subsection{Leveraging Adaptive Routing Notifications}
 \label{sec:implementation:arn}

Switches need store the information of packets responsible of creating congestion situations, such as the destination identifier, output port, virtual channel, etc., to keep track of congestion roots. This information is propagated to upstream switches using adaptive routing notification (ARN) messages, as described in Section \ref{sec:background:arn}.
ARN$+$AFI leverages this mechanism, and stores the information related to congestion roots in the ARN table \cite{patenteARN}, which composed of a set of entries, each one with the fields shown in Table \ref{tab:ARN}.


 \begin{table}[ht!]
 \renewcommand{\arraystretch}{1.3}
 \caption{ARN Table entry fields and size.}
 \label{tab:ARN}
 \centering
 \begin{tabular}{|p{1,4cm}|p{5,7cm}|p{0,6cm}|}
 \hline 
  Field Name & Description & Size (bits)\\
   \hline 
    \hline 
    Destination & Identifies the destination triggering the congestion detection. & 16\\
   \hline 
    Port & Port of the switch where congestion is detected or receives an ARN message.  & 8 \\
   \hline
    VC & Identifies the queue/VC triggering the congestion detection.  & 8 \\   
   \hline 
    ARNid & Identifies uniquely the congestion root.  & 32 \\
    \hline 
    congestion RootInfo & Information dependent on the network topology that triggers the notification consumption. In Fat trees, this information is the stage where the congestion root is detected.  & 8 \\
    \hline 
    Consumed & ARN consumption state.  & 1 \\
   \hline  
 \end{tabular}
\end{table}

When ARN$+$AFI detects a congestion root at a given output port, it creates an entry in the ARN table containing the information about the packet responsible of congestion, i.e., the packet on top of the VOQ that exceeded the HCDTh (see Section \ref{sec:implementation:detector}). The function $congestionRootDetected()$ shown in Pseudo-code \ref{algorithm:createARN} is the responsible of this process.







%
%

%
%

\algrenewcommand\algorithmicindent{1em}%
\algnewcommand\algorithmicforeach{\textbf{for each}}
\algdef{S}[FOR]{ForEach}[1]{\algorithmicforeach\ #1\ \algorithmicdo}
\algnewcommand\algorithmicswitch{\textbf{switch}}
\algnewcommand\algorithmiccase{\textbf{case}}
\algdef{SE}[SWITCH]{Switch}{EndSwitch}[1]{\algorithmicswitch\ #1\ \algorithmicdo}{\algorithmicend\ \algorithmicswitch}%
\algdef{SE}[CASE]{Case}{EndCase}[1]{\algorithmiccase\ #1}{\algorithmicend\ \algorithmiccase}%
\algtext*{EndSwitch}%
\algtext*{EndCase}%

\newcommand{\vars}{\texttt}
\newcommand{\func}{\textsf}
\let\oldReturn\Return
\renewcommand{\Return}{\State\oldReturn}
\newcommand\NoDo{\renewcommand\algorithmicdo{}}
\newcommand\ReDo{\renewcommand\algorithmicdo{\textbf{do}}}
\newcommand\NoThen{\renewcommand\algorithmicthen{}}
\newcommand\ReThen{\renewcommand\algorithmicthen{\textbf{then}}}

  \begin{algorithm}[!htb]
   \scriptsize
   \caption{Filling the information of an ARN when detecting a congestion root. This function calls \texttt{processARN()}.}
  \label{algorithm:createARN}
  
    \begin{algorithmic}[1]
     \scriptsize
\ReThen  

\Function{congestionRootDetected}{Pckt, ARNid}
    \If{\vars{isInUpwardPhase(Pckt.oPort)}}
        \State{\vars{ARNInfo.congestionRootInfo} $\leftarrow$ \vars{Stage}}
    \Else
        \State{\vars{ARNInfo.congestionRootInfo} $\leftarrow$ \vars{Stage - 1}}
    \EndIf
    \State{\vars{ARNInfo.Destination} $\leftarrow$ \vars{Pckt.Destination}}
    \State{\vars{ARNInfo.VC} $\leftarrow$ \vars{Pckt.VC}}
    \State{\vars{ARNInfo.Port} $\leftarrow$ \vars{Pckt.oPort}}
    \State{\vars{ARNInfo.ARNid} $\leftarrow$ \vars{ARNid}}
    \State{\vars{processARN(ARNInfo, Pckt.AdaptedBit)}}
\EndFunction
\end{algorithmic}
\end{algorithm}

Basically, this function is in charge of filling the fields described above for the ARN message, using the $ARNInfo$ structure and the $Pckt$ received as parameters.
Note that the value of the $congestionRootInfo$ field depends on whether the routing is in the upward or downward phase. Based on this information, the ARN can be consumed in upstream switches whenever the switch stage is lower than the $congestionRootInfo$ value (see Section \ref{sec:background:arn}).
Once the ARN message fields are filled, Pseudo-code \ref{algorithm:createARN} calls $processARN()$.

  \begin{algorithm}[!htb]
   \scriptsize
   \caption{Processing incoming ARN messages.  This function calls \texttt{selectAlternativeOport()}.}
  \label{algorithm:handleARN}
  
    \begin{algorithmic}[1]
     \scriptsize
\ReThen  


\Function{processARN}{ARNInfo, AdaptedBit}
    \If{\vars{existsARNTableEntry(ARNInfo)}}
        \State {\vars{resetARNTableEntryTimer(ARNInfo)}}
     \Else
        \State{\vars{removeARNEntry(} \par
        \hskip\algorithmicindent 
        \vars{ARNInfo.Destination, ARNInfo.Port, ARNInfo.VC)}}
        
        \If{\vars{AdaptedBit == FALSE}}
            \If{\vars{Stage == ARNInfo.congestionRootInfo}}
                \State{\vars{selectAlternativeOport(}\par
                \hskip\algorithmicindent 
                \hskip\algorithmicindent
                \hskip\algorithmicindent
                \vars{ARNInfo.Destination, ARNInfo.VC)}}
                \State{\vars{ARNInfo.Consumed} $\leftarrow$ \vars{TRUE}}
            \Else
                \State{\vars{ARNInfo.Consumed} $\leftarrow$ \vars{FALSE}}
            \EndIf
            \State{\vars{addARNTableEntry(ARNInfo)}}            
        \EndIf
    \ReThen
    \EndIf
\EndFunction

\end{algorithmic}
\end{algorithm}

Pseudo-code \ref{algorithm:handleARN} shows the $processARN()$ function, which is called when a new congestion root is detected (see Pseudo-code \ref{algorithm:createARN}) and when an ARN message is received at some switch (see Pseudo-code \ref{algorithm:routing}).
This function receives an $ARNInfo$ and a value for the $AdapedBit$ of the packet responsible when congestion is detected.
$ProcessARN()$ resets the ARN entry timer if the $ARNInfo$ already exists in the ARN table, otherwise it removes old ARN table entries matching with the $Destination$, $Port$ and $VC$ of $ARNInfo$. Note that congestion roots may change their location so that the $ARNid$ will change, and the table entries pointing to congestion roots that have moved are removed as soon as possible. 
The $processARN()$ function also decides if there is an alternative route for the $Destination, VC$ pair in the $ARNInfo$ (and selects an alternative output port for the ARN entry), and if the switch is placed in the same stage that the $ARNInfo.congestionRootInfo$, then the ARN entry is set as consumed (i.e., it will not be propagated).
The last step is to create a new ARN entry in the table, whenever the $ARNInfo$ is not already included in the table.

Pseudo-code \ref{algorithm:alternativePort} shows the functionality to calculate an alternative output port for a $Destination, VC$ pair. It also computes the $VC$ in the next switch in which the packet will be stored. Note that ARN$+$AFI always select the $AFC$ for storing packets in the next switch, since the $selectAlternativeOport$ function is called when an ARN message is consumed and it is required to select an alternative route. In this way, deflected traffic flows are stored in the AFC and congestion spreading is reduced.

  \begin{algorithm}[htb]
   \scriptsize
   \caption{Computing an alternative output port for a packet addressed to a congesting point, given $Destination$ and $VC$.}
  \label{algorithm:alternativePort}

%
%

    \begin{algorithmic}[1]
     \scriptsize
\ReThen  
\Function{selectAlternativeOport}{Destination, VC}
    \If{\vars{usingAFI}}
        \State{\vars{VCinNextSwitch} $\leftarrow$ \vars{AFC}}
    \Else
        \State{\vars{VCinNextSwitch} $\leftarrow$ \vars{VC}}
    \EndIf
    \State{\vars{AlternativePort} $\leftarrow$ \par
           \vars{oportMaxNumberOfAvailableCredits(VCinNextSwitch)}}
    \State{\vars{setAlternativeOportInRoutingTable(}\par
           \vars{Destination, AlterativePort)}}
\EndFunction

\end{algorithmic}
\end{algorithm}

Pseudo-code \ref{algorithm:routing} shows the input port controller function, which handles incoming packets to switches (i.e., data and ARN packets). Moreover, it is also responsible for sending ARN messages to upstream switches, when congesting packets are received.

  \begin{algorithm}[htb]
   \scriptsize
   \caption{Input Port Controller behavior when receiving incoming packets (either data packets or ARNs). This method either processes an ARN (\texttt{processARN()}) or generates a new one if the data packet matches the ARN Table (\texttt{sendARN}), which is sent upstream to a neighboring switch.}
  \label{algorithm:routing}
  
    \begin{algorithmic}[1]
     \scriptsize
\ReThen  


\Function{inputPortController}{Pckt}
    \If{\vars{Pckt.type == ARN}}
        \State{\vars{processARN (Pckt, FALSE)}}
    \ElsIf {\vars{Pckt.AdaptedBit == FALSE}}
        \State{\vars{ARNInfo.Destination} $\leftarrow$ \vars{Pckt.Destination}}
        \State{\vars{ARNInfo.VC} $\leftarrow$ \vars{Pckt.VC}}
        \If{\vars{existsARNTableConsumedEntry (\&ARNInfo)}}
            \State{\vars{Pckt.oPort}$ \leftarrow$ \vars{getAlternativeOport(Pckt.Destination)}}
            \If{\vars{usingAFI}}
                \State{\vars{Pckt.AdaptedBit} $\leftarrow$ \vars{TRUE}}
            \EndIf
            \Else
                \If {\vars{existsARNTableEntry (\&ARNInfo)}}
                    \State{\vars{sendARN(inputPortID, ARNInfo)}}
                \EndIf
            \State{\vars{Pckt.oPort} $\leftarrow$ \vars{getDeterministicOport(Pckt.Destination)}}
        \EndIf            
    \Else{
        \vars{Pckt.oPort} $\leftarrow$ \vars{getDeterministicOport(Pckt.Destination)}}
    \EndIf
\EndFunction

\end{algorithmic}
\end{algorithm}

The input port controller calls the $processARN()$ function (see Algorithm \ref{algorithm:handleARN} when it receives a ARN message, otherwise it checks if a regular incoming packet matches any of the ARN Table entries marked as consumed. If so, it selects an alternative port and sets the adaptedBit when ARN$+$AFI is used.
However, if the packet matches an entry which is not consumed, ARN$+$AFI will send ARN message to upstream switches.
It is important to mention that packets with the ``adaptedBit'' set will preserve their original VC stored in their header. This is required for the ARN table updating purposes. Indeed, a marked packet may be moved to the AFC, being responsible of a congestion situation. In this case, an ARN will be generated containing that packet's original VC, instead of the AFC, and the receiving switches will update the ARN Table entry using the original VC of the congesting packet. In this way, the entries of the ARN table keep updated to identify congestion trees lasting for longer. Note that these marked packets once they select an alternative port, they will use deterministic routing in the next switches.



\subsection{Implementation details}
\label{sec:implementation:details}

Our proposal can be implemented in available interconnect technologies without requiring significant modifications in the switch architecture hardware.
The first thing that we need is that the network implements the ARN mechanism or any similar alternative. For instance, the InfiniBand HDR networks have support for the ARN mechanism and congestion detection, so that these networks would benefit of our proposal. According to the Mellanox/NVIDIA documentation \cite{MellanoxConfigureAR}, adaptive routing (AR) enables the switch to select the output port to route a packet based on the port's load.
The second thing that the network needs is to offer support for static queuing schemes (SQS). Note that most of the commercial high-performance network technologies implement VCs, and they also have flexibility to configure the switch buffers to use static queuing schemes (SQS) and the associated tables that permit the mapping of traffic classes (TCs) or service levels (SLs) to the available VCs (see \figurename~\ref{fig_sw_architecture}). For instance, we demonstrated in the past that SQSs are suitable for InfiniBand-based networks \cite{ESCUDEROSAHUQUILLO201835} 
In these tables, called SL-to-VL in IBA-based networks, we need the hardware support to look at the ``adapted'' bit at the packet header, so that the adapted flow channel (AFC) can be selected in the following switch to store those packets being routed through alternative routes (see Section \ref{sec:implementation:architecture}). This functionality is simple because the switches look at the packet header information to obtain the TC or SL value when a packet needs to be forwarded to the next switch. Therefore, we only need to incorporate the ``adapted'' bit value in the result of the query to get the TC or SL information from the packet.
Moreover, we also need that the packet header format is extended to include the ``adapted'' bit, which our proposal needs to check at the SL-to-VL tables to set the VC (either a regular VC or the AFC) where that packet will be stored in the next switch.

In summary, the required modifications in the network architecture are the special bit in the packet header to set when a packet is ``adapted'' by the AFI mechanism, and the functionality at the SL-to-VL tables to look the packet header bit and select the AFC in the next switch when the bit is set to one. Note that these modifications do not require significant modifications in the network architecture, so that we consider that they are easily assumable by the network manufacturers.

\begin{figure*}[thb!]
\centering
\includegraphics[width=1\columnwidth]{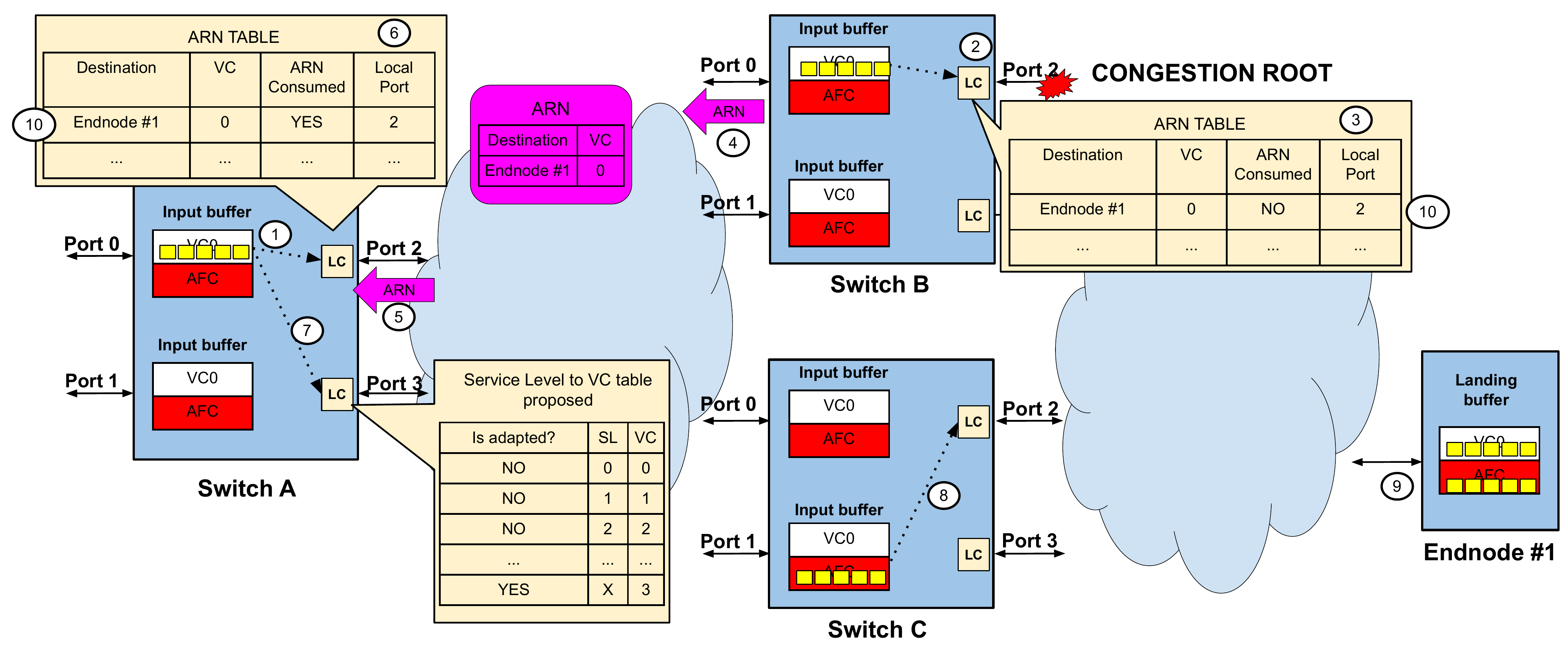}
\caption{Operation example of our proposal in a network portion. $4$ VCs are used (VC 3 being the AFC).}
\label{fig_example}
\end{figure*}

\subsection{Operation Example}
 \label{sec:implementation:operations}
 
\figurename~\ref{fig_example} shows an operation example of our proposal when congestion appears in a network portion
including three switches (A, B and C) and the destination endnode \#1.
At each switch input port, we assume that two VCs are used; the first VC is used for regular traffic and the second one (in red) is the AFC. For the sake of clarity, we have omitted the VOQs in the example.
We assume an indeterminate number of switches between Switch A and both switches B and C, and between these switches and the destination endnode \#1, illustrated by means of clouds.
In this example, we assume a transmission from input port \#0 at switch A (Event \#1) to the destination endnode \#1, where packets follow a deterministic route through Switch B.

At a given time instant, the congestion detector in switch B identifies a congestion root in the output port \#2 (Event \#2), and determines that the VC0 at the input port \#0 stores the packet responsible of this situation.
Note that to identify that packet, the switch B needs to look at the VOQ 
\#2 in input port \#0.
As the packet responsible of congestion is addressed to destination endnode \#1, then this destination is considered as congested.
In this point, switch B creates an entry in the ARN table (Event \#3), containing the destination endnode \#1, the output port \#2, the VC \#0, the ARNid, and the \emph{CongestionRootInfo} (i.e., the stage in the Fat-Tree in which this congestion is detected).

As the ARN table can be queried from any input port in the switch, each time an input port wants to send a packet matching the ARN entry formerly created for Switch B (i.e., endnode \#1, local output port \#2, VC \#0, etc.), then an ARN is sent back to the previous switch (Event \#4), containing the destination endnode \#1 and the VC \#0.
This ARN is propagated backwards throughout the network, whenever there are switches that do not consume it.
In the example, we assume that this ARN is not consumed until it reaches Switch A (Event \#5).
Note that if an ARN is not consumed when it arrives to a switch, that switch still stores in the ARN table the information contained in that ARN.
When Switch A receives the ARN, it checks if the ARN can be consumed by looking at the \emph{CongestionRootInfo} field. 
In this example, we assume that Switch A satisfies this condition, 
so that, it consumes the ARN (Event \#6) and creates the corresponding ARN table entry.
From that moment, all the packets matching that entry will be routed through an alternative path.
For instance, all the packets stored in the VC0 at input port $0$, which are addressed to destination endnode \#1 will be marked as ``adapted'' and routed through the alternative route that traverses the output port $3$ (Event \#7).
In the next switch, these packets will be stored in the AFC.
Later, these ``adapted'' packets will reach Switch C, and they will be stored in the AFC at input port 1 (Event \#8).
As mentioned above, the packets stored in the AFC are routed using deterministic routing, and when they arrive to other switches they will be stored in the AFC.
Finally, the ``adapted'' packets arrive to the destination endnode \#1 through the AFC (Event \#9), while the non-adapted ones arrive through the VC0 using the deterministic path.
When the congestion situation ends on switch B, it stops updating the table (Event \#10). Since they are not updated, the entries in the ARN tables will disappear after a time period (time-to-live).

\section{Performance Evaluation}
 \label{sec:evaluation}


\subsection{Simulation model}
 \label{sec:evaluation:model}
For the experiments carried out in this section, we use SAURON \cite{Yebenes13PDP}, a custom-made event-driven simulator based on the OMNeT++ framework \cite{OMNeTweb}.
We have extended the SAURON tool to include the support for Adaptive Routing Notifications (ARNs) \cite{patenteARN}, and to model our proposal based on the details shown Section \ref{sec:implementation}.
We have also combined our proposal with static queuing schemes (SQSs), as we describe below.
Regarding the network configuration, we have modeled the Fat-tree networks shown in Table \ref{tab:networks}.
Specifically, we have modeled two 3-stage Real-Life Fat-Tree (RLFT) networks, each one using switches with different port count (i.e., 12 and 24).

 \begin{table}[ht!]
 \renewcommand{\arraystretch}{1.3}
 \caption{Evaluated RLFT network configurations}
 \label{tab:networks}
 \centering
 \begin{tabular}{|c|r|r|r|r|}
 \hline 
  \# & Endnodes & Port Count & Stages  & Switches \\
   \hline 
    \hline 
  1	& 432 & 12 & 3  & 180 \\
   \hline 
  2 & 3456 & 24 & 3  & 720 \\
   \hline 
 \end{tabular}
\end{table}

We have modeled input-queued (IQ) switches, i.e., switches with buffers only present at input ports (see Section \ref{sec:implementation:architecture}).  The size of each input-port buffer is $344$KB and the packet MTU is $4$KB, therefore $84$ packets can be stored at every input-port. Each input-port buffer is divided into the VCs required for the SQS used, so the size of each VC depends on the number of required VCs per port. Note that physical links are divided into as many logical (i.e., virtual) links as VCs are available. 
The flow-control policy assumed in our model is credit-based in order to guarantee ``lossless'' network behavior, meaning that packet dropping is not allowed.

We have also modeled virtual output queues (VOQs), which allows the access to the switch crossbar to be demultiplexed, i.e., packets from the same input buffer requesting different output ports can be forwarded at the same time.
Thus, we prevent the low-order HoL-blocking appearing at input buffers.

Each one of the endnodes has a network interface (a.k.a., host channel adapter or HCA using the InfiniBand terminology).
HCAs connect the endnodes to the network, and split into packets the messages generated by the applications.
For the network links, we assume serial full-duplex links of $5$-meter length with $100$ Gbps of link bandwidth, and $30$ nanoseconds of link propagation delay (i.e., $6$ns/meter).
In the following, we describe the techniques modeled in the simulator in order to evaluate our proposal.

    \textbf{$D$-mod-$K$}.This deterministic routing is used to measure performance when the routing algorithm does not offer multi-path \cite{Gomez07ipdps, Zahavi10cpe}.
    
    \textbf{Oblivious}. This routing algorithm (also known as multi-path routing) selects randomly an output port from those than can be chosen to route minimally a packet, without considering traffic conditions.
    
    \textbf{Adaptive-Th}. We have also modeled threshold-based adaptive routing algorithms for Fat-Trees, described in Section \ref{sec:background:fat-trees}. We assume that it is triggered when a buffer occupancy exceeds 75\% of its capacity. We tuned this value to allow routing algorithm to adapt only when a congestion situation is strong and lasting.
    
    \textbf{Adaptive-Th + AFI}. This configuration combines the previous routing algorithm with AFI \cite{rocherCCgrid}, to measure the congestion spreading reduction.
    
    \textbf{ARN mechanism}. We have modeled this mechanism using the information made publicly available  by Mellanox/NVIDIA (i.e., the main manufacturers of InfiniBand-based hardware) CITAR [Hipineb17], [NVIDIA-quantum].
    Also, we assume that the ARN mechanism uses the congestion detector described in Section \ref{sec:implementation:detector}.
    
     \textbf{ARN + AFI}. As mentioned in Section \ref{sec:implementation}, our proposal combines the use of AFI and ARNs. The congestion detector parameters have been set as for the ARN mechanism.
     
Specifically, for \emph{ARN} and \emph{ARN+AFI} we have configured the congestion detector according to the following parameters. HCDTh is set to 81\% of the buffer occupancy and LCDTh is set to 63\%. On the other hand, FCTh is set to 78 \% of the free credits in the next buffer, over this threshold the congested port is considered a congestion root. To declare a root, the root situation must persist for some time. This period is called CRT and is set to $5$ ms. We have tuned thoroughly these parameters by means of simulations in order to use the most appropriate values. In more detail, HCDTh and LCDTh have been tuned based on the experiments performed by Gran et al. \cite{GranCCgrid}. FCTh has been tuned assuming that several packets may be on the fly on their way from one switch to the following one. Finally, CRT has been tuned considering the fastest reaction of the detector when congestion trees are in steady state.


For all these techniques, we assume the use of $1$VC, except when AFI is used, which needs the AFC to store adapted packets.
Moreover, as our proposal can be combined with static queuing schemes (SQSs) in order to achieve an extra HoL blocking reduction, we have modeled the techniques described above combined with several SQS proposed for Fat-Trees, such as DBBM, vFtree or Flow2SL (see Section \ref{sec:background:congestion_management}).
We assume the SQSs use $3$VCs, plus the AFC if they are combined with AFI.

Regarding the performance metrics, we have measured the network efficiency versus simulation time (set to $120$ms) when synthetic traffic is used. Note the network efficiency is the average throughput (i.e., the amount of packets that the network delivers per time unit) normalized with respect to one.
Also, we have measured the execution time when trace-based traffic is generated in the network, which shows the time spent by the network in processing the messages recorded in the trace file. 
Note that this metric shows the network congestion impact when running the mentioned application.

\subsection{Traffic Model}
 \label{sec:evaluation:trafic-model}

We have modeled two traffic patterns which are representative of those generated in the network by the workloads used in HPC systems and Data-centers. Specifically, we have fed our simulation tool with synthetic and trace-based traffic patterns, as we describe below.

\subsubsection{Synthetic traffic}
\label{sec:evaluation:traffic-model:synthetic}

The synthetic traffic is commonly used in network simulators in order to model specific communication patterns in the network and analyze how these patterns affect the network performance.
For this study we have modeled three communication patterns: uniform, single incast and four incast.

First, in the \textbf{uniform (random) traffic}, all the endnodes in the network generate packets addressed to a uniform distribution of destinations, which are selected randomly.
This traffic pattern is also known as \emph{all-to-all}.
The generation rate at source endnodes is 100\%, which means that links operate at $100$ Gbps.
This traffic pattern (\emph{RND}) is intended to measure the network performance when the load generation rate is high and congestion situations are small and temporary, due to the random traffic distribution.

\begin{figure}[!htb]
\centering
\begin{subfigure}{0.48\columnwidth}
\centering
\includegraphics[width=.95\columnwidth]{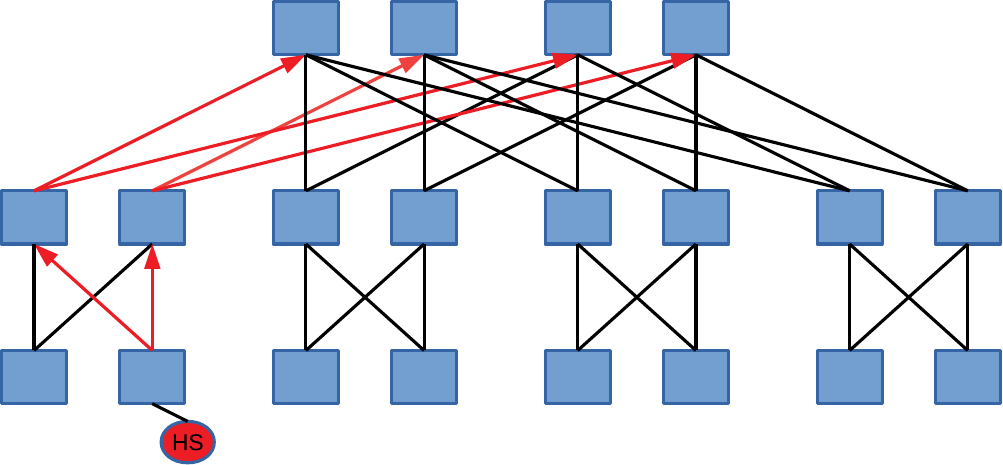}
\vspace{-.2cm}
\caption{A single Hot-Spot.}
\vspace{.3cm}
\label{fig_traffic_HS1}
\end{subfigure}
\begin{subfigure}{0.48\columnwidth}
\centering
\includegraphics[width=.95\columnwidth]{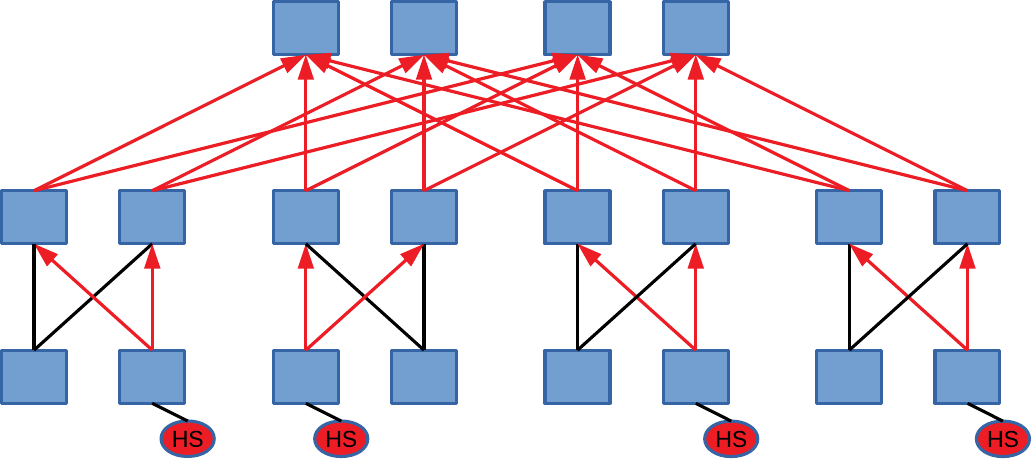}
\caption{4 Hot-Spots.}
\label{fig_traffic_HS4}
\end{subfigure}

\caption{Congestion scenarios in a $3$-stage RLFT. Red arrows show the paths towards where congestion trees grow.}
\label{fig_traffic_scenarios}
\end{figure}

Another synthetic traffic pattern used in this study is the \textbf{single incast congestion scenario}. \figurename~\ref{fig_traffic_HS1} shows a single incast (or congestion root) generated near the endnode ``HS''.
This situation, is generated because several endnodes inject traffic in the network following a many-to-one communication pattern.
The red arrows show the direction to where the congestion tree grows.
Note that, when adaptive routing is used, the traffic flows can be spread through out the available routes reaching the endnode ``HS''.
We have modeled two single incast scenarios: \emph{H10} and \emph{H25}. For \emph{H10}, 10\% of the endnodes generate traffic addressed to a single endnode.
They start to inject traffic after a warm-up period of $3$ms, and injection lasts for $90$ms.  The simulation time is set to $120$ms. The remainder endnodes (90\%) generate random traffic  during all the simulation time.
For \emph{H25}, 25\% of endnodes generate traffic addressed to a single destination, while the remainder ones (75\%) generate traffic addressed to a uniform distribution of random endnodes.
In both cases, the generation rate at source endnodes is 100\%. 
The hot-spot is at endnode $4$ for both network configurations \#1 and \#2 (see Table \ref{tab:networks}).

Finally, a corner-case situation that may spoil the network performance is when several incast situations (i.e., more than one congestion trees) co-exist in the network.
We have modeled a \textbf{four incast congestion scenario}. \figurename~\ref{fig_traffic_HS4} shows four congestion trees generated by several many-to-one traffic patterns.
For this traffic pattern, we have also modeled two scenarios: \emph{H10-4} and \emph{H25-4}. For \emph{H10-4}, 10\% of the endnodes generate traffic addressed to four endnodes, while the remainder endnodes (90\%) generate random traffic.
For \emph{H25-4}, 25\% of the endnodes generate traffic addressed to four endnodes, and the remainder ones (75\%) generate random traffic.
The hot-spots are also generated after the warm-up period, and the links operate at full speed during $90$ ms.
For network configuration \#1 the hot spots are at endnodes $4$, $120$, $244$, and $431$ while for configuration \#2 at endnodes $0$, $889$, $1772$, and $3454$.

\subsubsection{MPI-based traces}
\label{sec:evaluation:traffic-model:traces}

We have modeled realistic communication operations in the network using the VEF traces framework \cite{Andujar16JSC}. 
This framework allows to capture the network communication of MPI-based applications run in a real cluster installation, and record this communication into traffic traces, used to fed the network simulator.
As the size of these traces is not large enough to feed $3456$-node Fat-trees, we have evaluated these traffic patterns under $432$-node Fat-trees (see network configuration \#1 in Table \ref{tab:networks}). Specifically, we have used two MPI-based traces from the PTRANS test and the Inception-v3 application. PTRANS, included in the HPCC benchmark \cite{PTRANS}, is composed of $432$ MPI tasks. PTRANS exercises the nodes communication of  heavily, based on a realistic problem where pairs of processors communicate with each other simultaneously.
Inception-V3 is a convolutional neural network (CNN) designed by Google and used for image classification \cite{inception}. It is composed of $512$ MPI tasks which perform the training phase of the CNN.

In order to measure the congestion impact on these applications, we have also modeled a situation where an incast scenario suddenly appears $3$ms after the application is running.
Note that an incast situation may appear in a HPC cluster or Data-center running MPI-based applications when one of them, such as a deep-learning application accessing to the data-sheet for training purposes, eventually and suddenly executes a many-to-one communication operation in the network (e.g., an MPI Gather collective operation), which generates an incast (or congestion) scenario \cite{Congdon21Nendica}.
More precisely, we have generated a congestion tree based on the \emph{HS10} model during $100$ ms for PTRANS and $400$ ms for Inception-v3, as the trace size for both applications is different. Note that endnodes generating communication operations for PTRANS or Inception-v3 are not intended to generate congested traffic. However, the congestion tree will delay their execution time if no congestion management is used, as we discuss later.
  
\begin{figure*}[!htb]
\begin{subfigure}{1\textwidth}
 \centering 
\includegraphics[width=1\textwidth]
{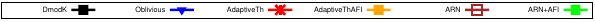}
\label{leyenda2}
\end{subfigure}

\begin{subfigure}{0.245\textwidth}
 \centering 
\includegraphics[width=1\textwidth]
{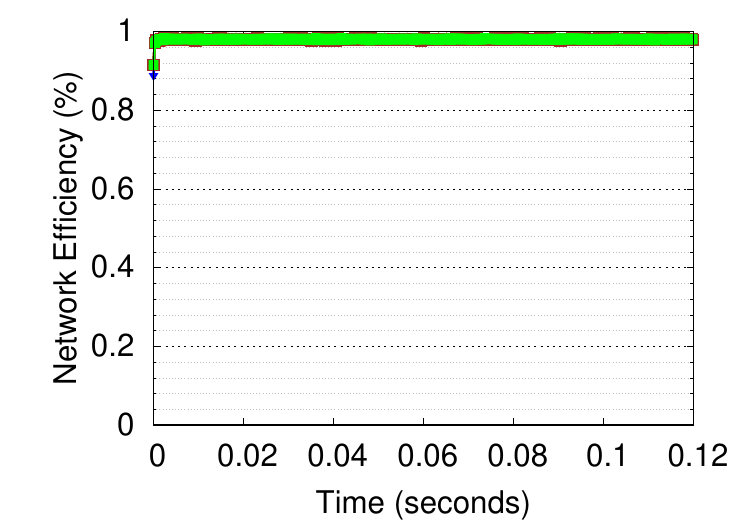}
\caption{RND - 1Q}
\label{fig_RLFT_RD_1q_T04_100_432}
\end{subfigure}
\begin{subfigure}{0.245\textwidth}
 \centering 
\includegraphics[width=1\textwidth]
{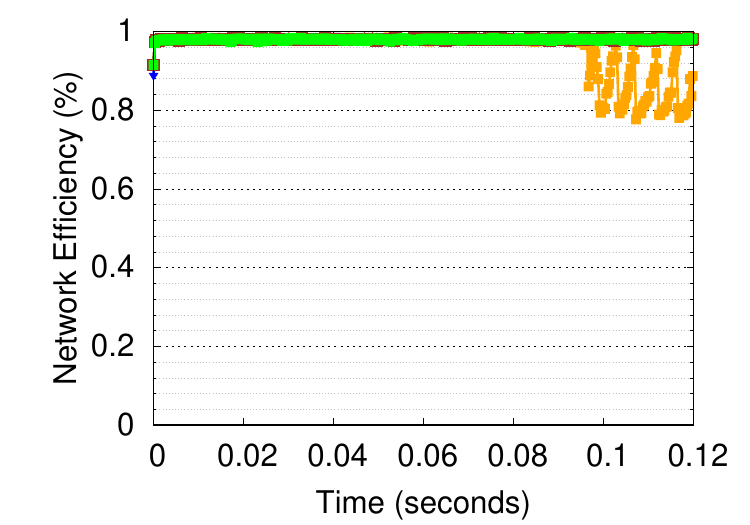}
\caption{RND - DBBM}
\label{fig_RLFT_RD_dbbm3_T04_100_432}
\end{subfigure}
\begin{subfigure}{0.245\textwidth}
 \centering 
\includegraphics[width=1\textwidth]
{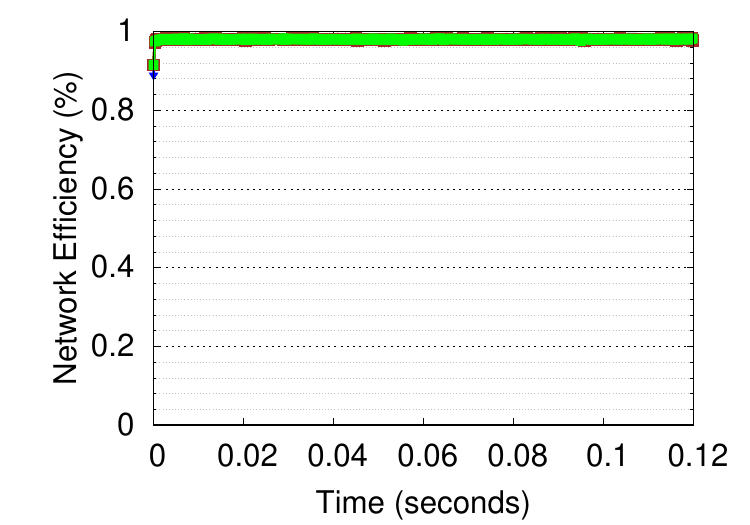}
\caption{RND - Vftree}
\label{fig_RLFT_RD_vftree3_T04_100_432}
\end{subfigure}
\begin{subfigure}{0.245\textwidth}
 \centering 
\includegraphics[width=1\textwidth]
{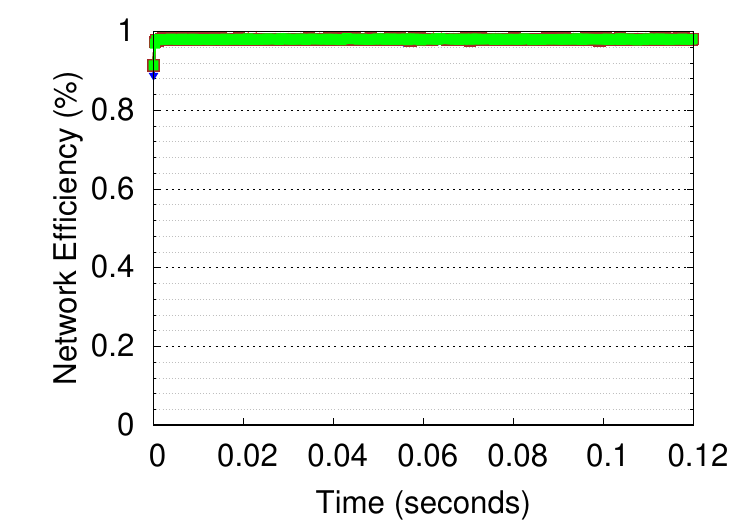}
\caption{RND - Flow2SL}
\label{fig_RLFT_RD_flow2sl3_T04_100_432}
\end{subfigure}

\begin{subfigure}{0.245\textwidth}
 \centering 
\includegraphics[width=1\textwidth]
{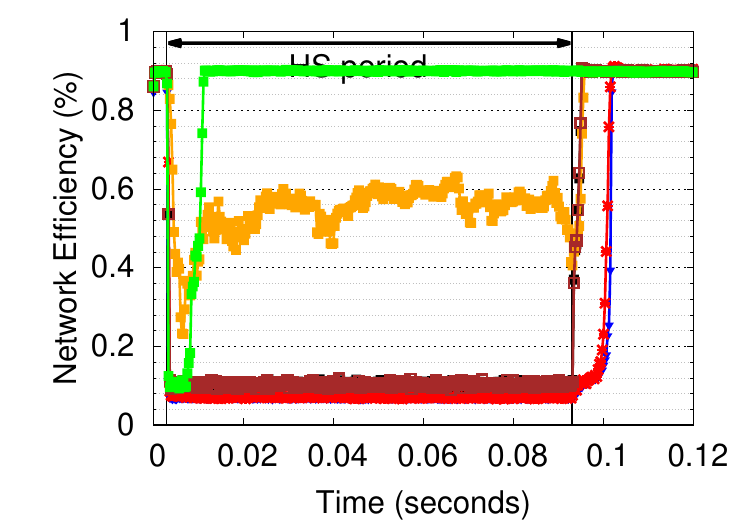}
\caption{H10- 1Q}
\label{fig_RLFT_HS10_1q_T04_100_432}
\end{subfigure}
\begin{subfigure}{0.245\textwidth}
 \centering 
\includegraphics[width=1\textwidth]
{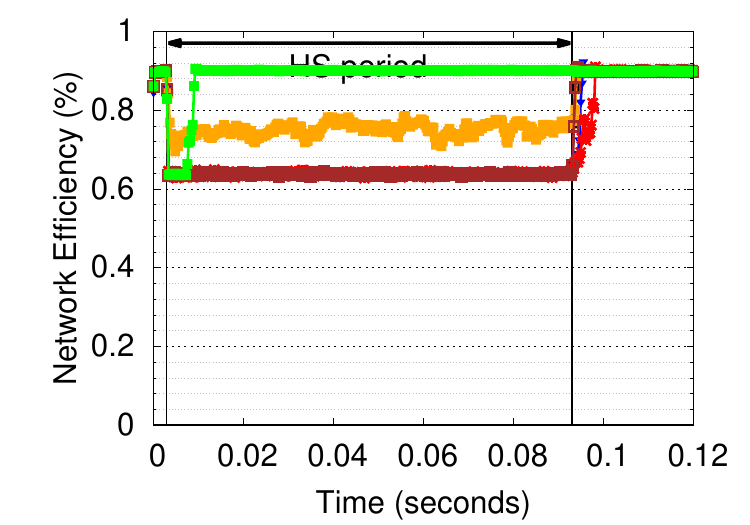}
\caption{H10- DBBM}
\label{fig_RLFT_HS10_dbbm3_T04_100_432}
\end{subfigure}
\begin{subfigure}{0.245\textwidth}
 \centering 
\includegraphics[width=1\textwidth]
{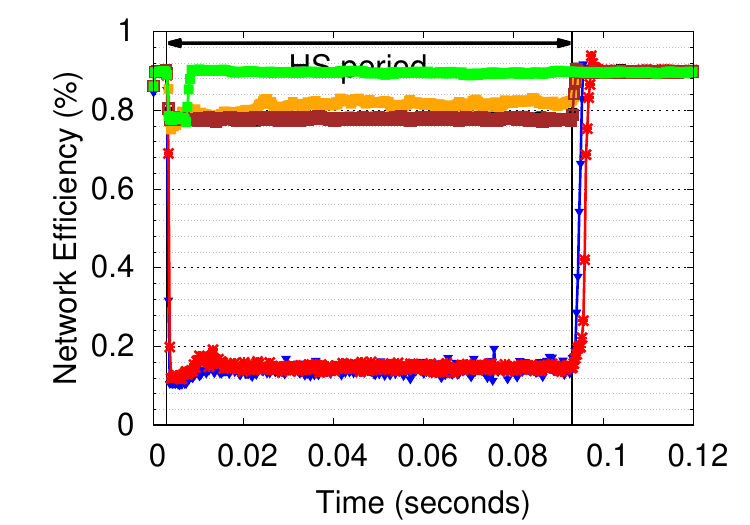}
\caption{H10- Vftree}
\label{fig_RLFT_HS10_vftree3_T04_100_432}
\end{subfigure}
\begin{subfigure}{0.245\textwidth}
 \centering 
\includegraphics[width=1\textwidth]
{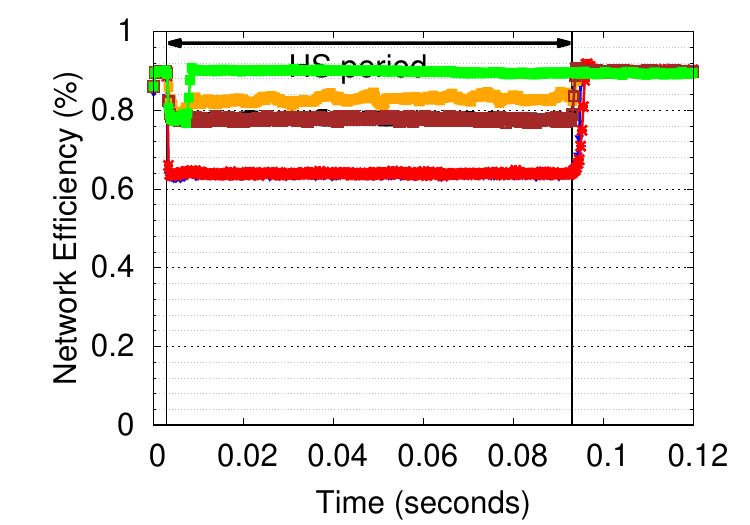}
\caption{H10- Flow2SL}
\label{fig_RLFT_HS10_flow2sl3_T04_100_432}
\end{subfigure}

\begin{subfigure}{0.245\textwidth}
 \centering 
\includegraphics[width=1\textwidth]
{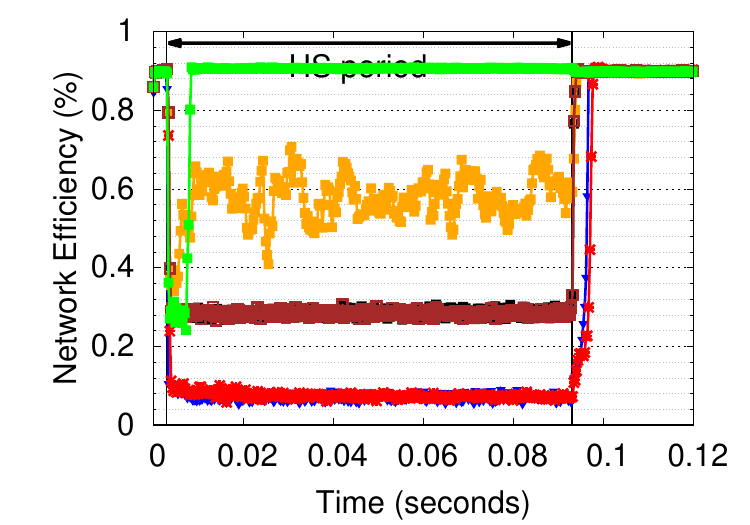}
\caption{H10-4 - 1Q}
\label{fig_RLFT_HS10-4_1q_T04_100_432}
\end{subfigure}
\begin{subfigure}{0.245\textwidth}
 \centering 
\includegraphics[width=1\textwidth]
{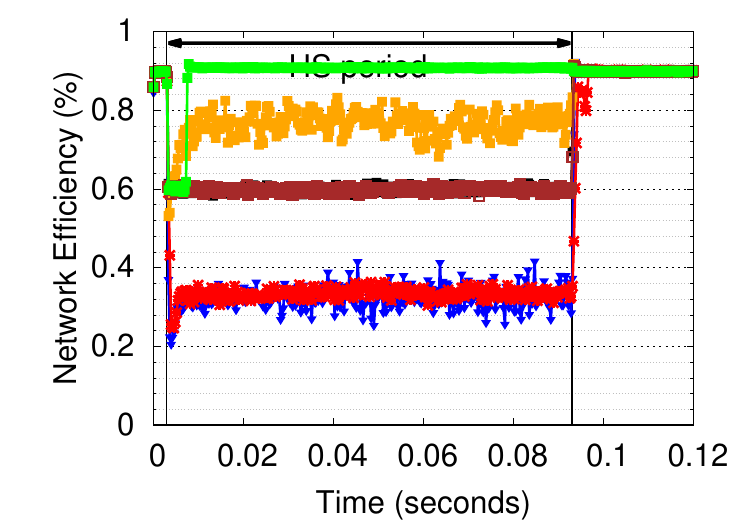}
\caption{H10-4 - DBBM}
\label{fig_RLFT_HS10-4_dbbm3_T04_100_432}
\end{subfigure}
\begin{subfigure}{0.245\textwidth}
 \centering 
\includegraphics[width=1\textwidth]
{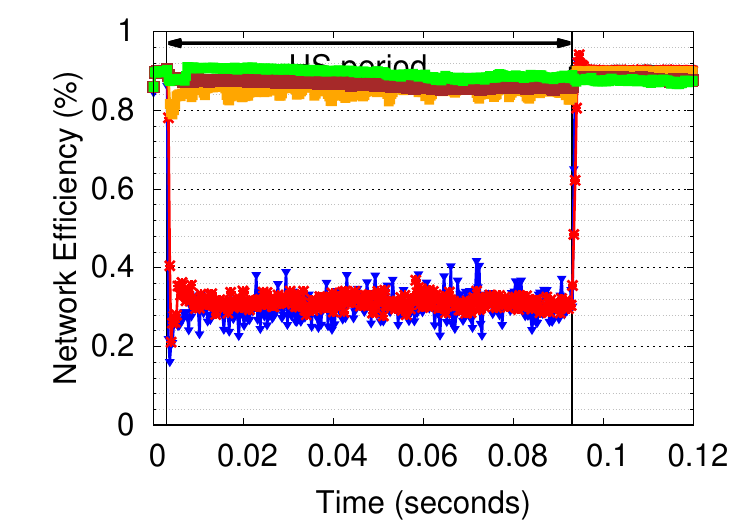}
\caption{H10-4 - Vftree}
\label{fig_RLFT_HS10-4_vftree3_T04_100_432}
\end{subfigure}
\begin{subfigure}{0.245\textwidth}
 \centering 
\includegraphics[width=1\textwidth]
{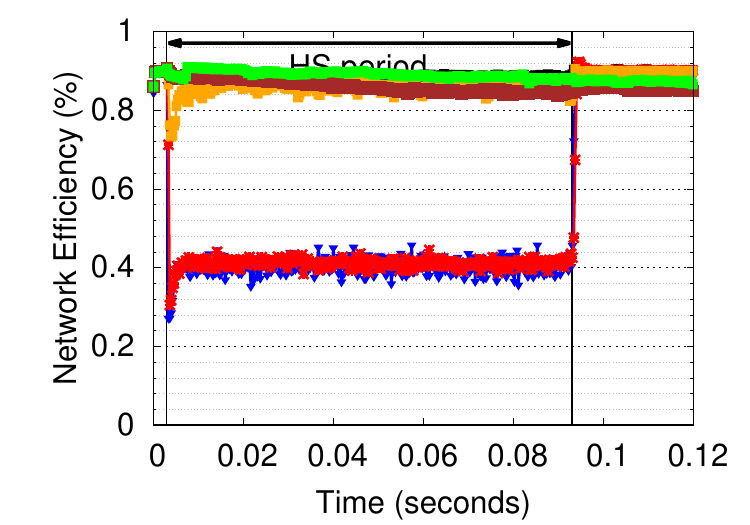}
\caption{H10-4 - Flow2SL}
\label{fig_RLFT_HS10-4_flow2sl3_T04_100_432}
\end{subfigure}

\begin{subfigure}{0.245\textwidth}
 \centering 
\includegraphics[width=1\textwidth]
{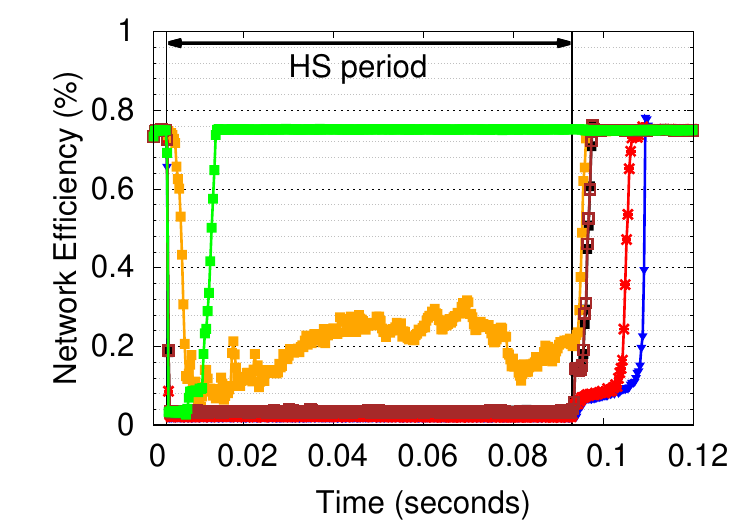}
\caption{H25  - 1Q}
\label{fig_RLFT_HS25_1q_T04_100_432}
\end{subfigure}
\begin{subfigure}{0.245\textwidth}
 \centering 
\includegraphics[width=1\textwidth]
{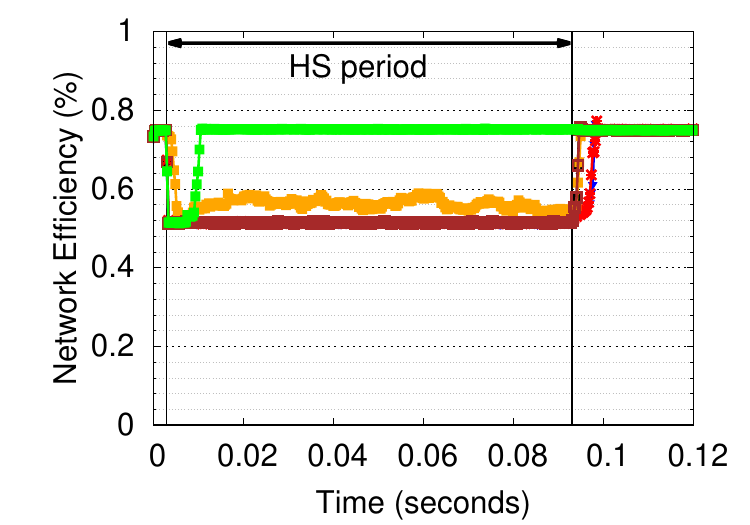}
\caption{H25  - DBBM}
\label{fig_RLFT_HS25_dbbm3_T04_100_432}
\end{subfigure}
\begin{subfigure}{0.245\textwidth}
 \centering 
\includegraphics[width=1\textwidth]
{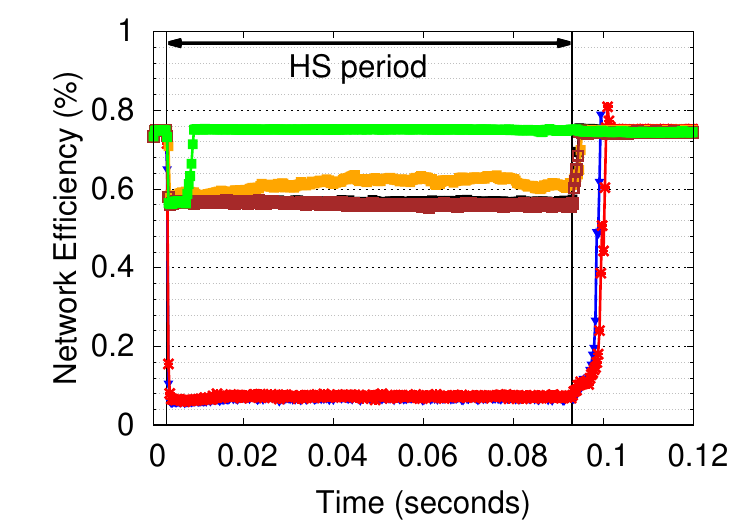}
\caption{H25  - Vftree}
\label{fig_RLFT_HS25_vftree3_T04_100_432}
\end{subfigure}
\begin{subfigure}{0.245\textwidth}
 \centering 
\includegraphics[width=1\textwidth]
{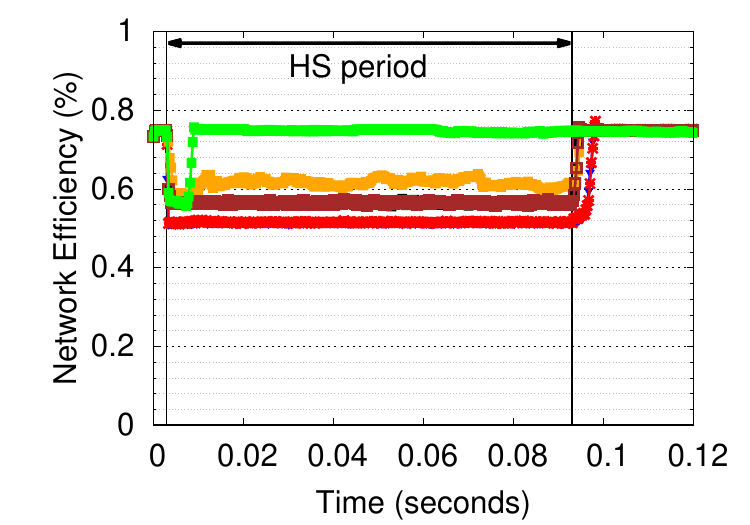}
\caption{H25  - Flow2SL}
\label{fig_RLFT_HS25_flow2sl3_T04_100_432}
\end{subfigure}

\begin{subfigure}{0.245\textwidth}
 \centering 
\includegraphics[width=1\textwidth]
{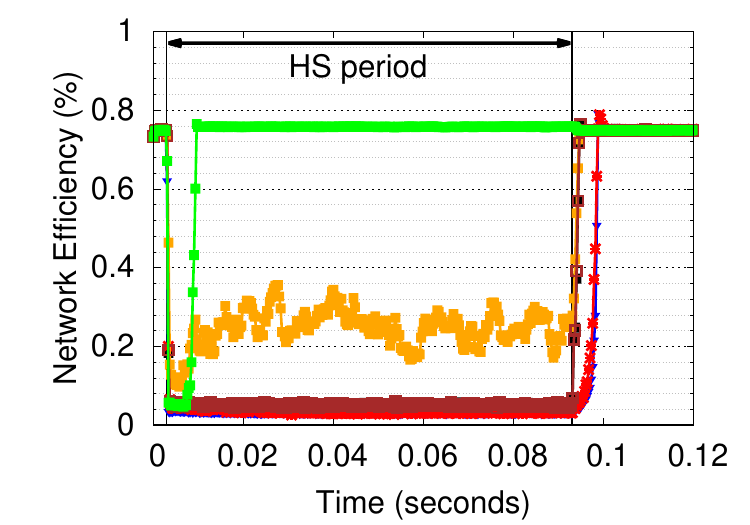}
\caption{H25-4 - 1Q}
\label{fig_RLFT_HS25-4_1q_T04_100_432}
\end{subfigure}
\begin{subfigure}{0.245\textwidth}
 \centering 
\includegraphics[width=1\textwidth]
{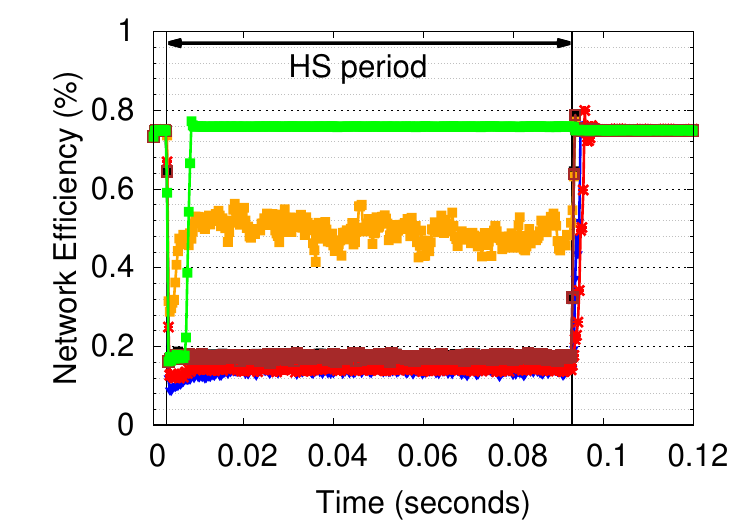}
\caption{H25-4 - DBBM}
\label{fig_RLFT_HS25-4_dbbm3_T04_100_432}
\end{subfigure}
\begin{subfigure}{0.245\textwidth}
 \centering 
\includegraphics[width=1\textwidth]
{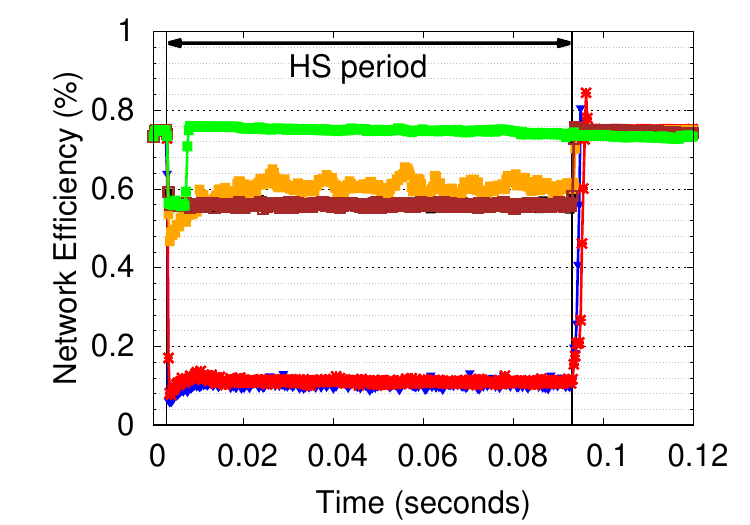}
\caption{H25-4 - Vftree}
\label{fig_RLFT_HS25-4_vftree3_T04_100_432}
\end{subfigure}
\begin{subfigure}{0.245\textwidth}
 \centering 
\includegraphics[width=1\textwidth]
{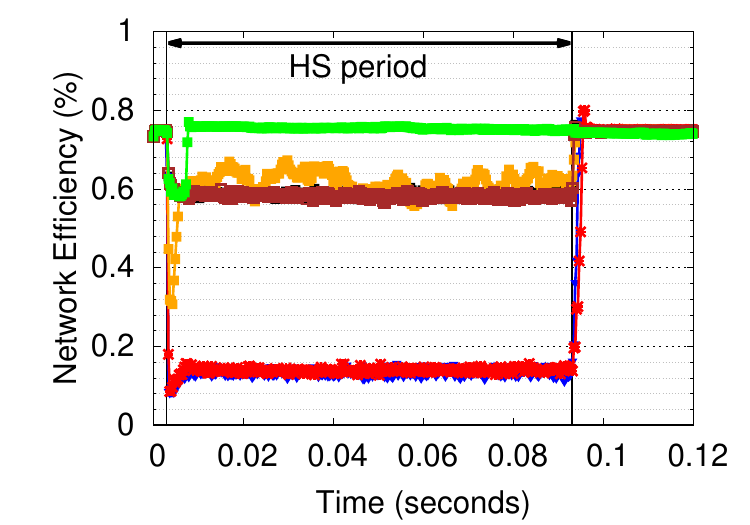}
\caption{H25-4 - Flow2SL}
\label{fig_RLFT_HS25-4_flow2sl3_T04_100_432}
\end{subfigure}

\caption{Network Efficiency versus Time for the modeled techniques in a $432$-node RLFT network (see configuration \#1 in Table \ref{tab:networks}), when traffic patterns described in Section \ref{sec:evaluation:trafic-model} are generated.}
\label{fig:RLFT_432_T04_100}
\end{figure*}

\subsection{Experiment results with synthetic traffic} 
 \label{sec:evaluation:sythetic-traffic}
 
\figurename~\ref{fig:RLFT_432_T04_100} shows the experiment results for the modeled techniques (see Section \ref{sec:evaluation:model}) in network configuration \#1 (see Table \ref{tab:networks}), i.e. a $432$-node RLFT, when synthetic traffic is generated (see Section \ref{sec:evaluation:trafic-model}).
We show the results for 1Q and SQSs (DBBM, Vftree and Flow2sl), so that $1$ VC is always used in the former case and $3$ VCs are used in the later except for AFI, which requires an additional VC for the AFC.
Figures \ref{fig_RLFT_RD_1q_T04_100_432} - \ref{fig_RLFT_RD_flow2sl3_T04_100_432} show the results for uniform traffic. Most of the techniques obtain a network efficiency close to $1$, since this traffic generates light congestion, and the high-order HoL-blocking barely appears.
Note that low-order HoL-blocking is natively prevented by the switch architecture, which uses VOQs (see Section \ref{sec:implementation:architecture}).
However, \emph{Adaptive-Th + AFI} experiences a significant performance degradation when DBBM is used due to its unfortunate combination of mapping policy of packets to VCs and routing when the network is saturated. 

\figurename~\ref{fig_RLFT_HS10_1q_T04_100_432} - \ref{fig_RLFT_HS10_flow2sl3_T04_100_432} show the obtained results when 10\% of the endnodes generate a single incast scenario (i.e., the \emph{HS10} traffic pattern). In this situation, the efficiency of all the techniques drops to $0.1$ of efficiency for the 1Q configuration (see \figurename~\ref{fig_RLFT_HS10_1q_T04_100_432}), when the hot-spot generation starts. Our proposal \emph{ARN+AFI} reacts quickly to this congestion situation and it is able to recover its efficiency level after less than $8$ms, since it prevents the congestion spreading generated by the ARN mechanism.
Note that the rest of the routing configurations also suffer from congestion spreading, and they are not able to recover their efficiency within a long period of time.
Note also that, when SQS are used, the performance dropping is lower than that obtained for 1Q, and \emph{ARN+AFI} also achieves the best results.
Moreover, \emph{$D$-mod-$K$} and \emph{ARN} routing schemes always have a similar behaviour,  since \emph{ARN} adapts only when it receives a notification and maintains over time the routing through the new port. However, as \emph{Oblivious} and \emph{AdaptiveTh} have a higher degree of adaptability, they are more likely to spread the congestion tree and always obtain the worst results.

\figurename~\ref{fig_RLFT_HS10-4_1q_T04_100_432} - \ref{fig_RLFT_HS10-4_flow2sl3_T04_100_432} show the obtained results when the \emph{HS10-4} traffic pattern is generated.
\emph{ARN+AFI} efficiently identifies the four congestion trees and isolates them in the AFC, reducing the congestion spreading. On the contrary, the rest of the techniques show a signif  icant performance dropping, even worse than that observed for the previous traffic pattern when SQSs are used. 
The main reason for that is the four congestion trees being generated, which makes it more complicated to isolate them, even though SQSs are used.
VFtree and Flow2SL combined with \emph{ARN+AFI}  reduce the performance dropping, compared to that observed for the 1Q and DBBM configurations.

\figurename~\ref{fig_RLFT_HS25_1q_T04_100_432} - \ref{fig_RLFT_HS25_flow2sl3_T04_100_432} and \figurename~\ref{fig_RLFT_HS25-4_1q_T04_100_432} - \ref{fig_RLFT_HS25-4_flow2sl3_T04_100_432}, show the simulation results when traffic patterns \emph{HS25} and \emph{HS25-4} are used, i.e., the number of sources generating congesting flows increases up to 25\%. Again, the \emph{ARN+AFI} technique obtains the best results, and similar conclusions can be drawn to those observed for \emph{HS10} and \emph{HS10-4}.

\begin{figure*}[!th]
\scriptsize
\begin{subfigure}{1\textwidth}
 \centering 
\includegraphics[width=1\textwidth]
{ ./leyenda.pdf}
\label{leyenda1}
\end{subfigure}

\begin{subfigure}{0.245\textwidth}
 \centering 
\includegraphics[width=1\textwidth]
{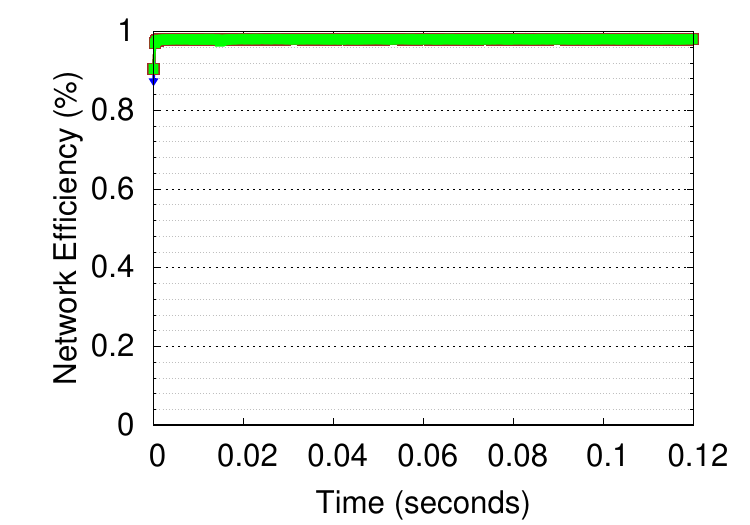}
\caption{RND - 1Q}
\label{fig_RLFT_RD_1q_T04_100}
\end{subfigure}
\begin{subfigure}{0.245\textwidth}
 \centering 
\includegraphics[width=1\textwidth]
{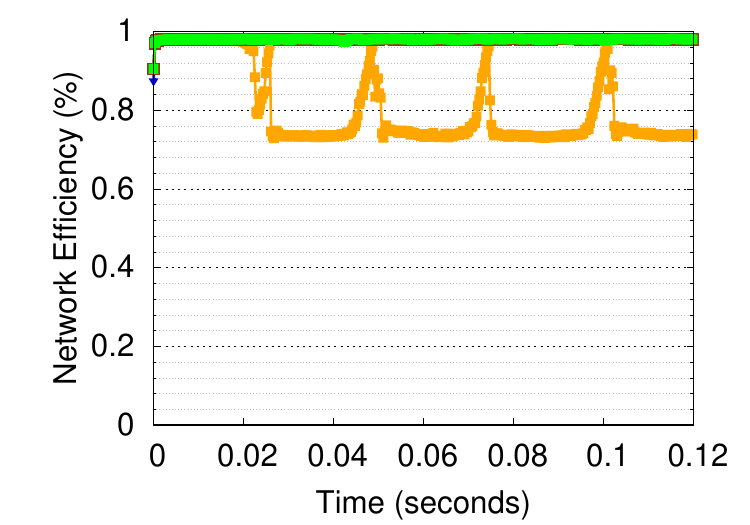}
\caption{RND - DBBM}
\label{fig_RLFT_RD_dbbm3_T04_100}
\end{subfigure}
\begin{subfigure}{0.245\textwidth}
 \centering 
\includegraphics[width=1\textwidth]
{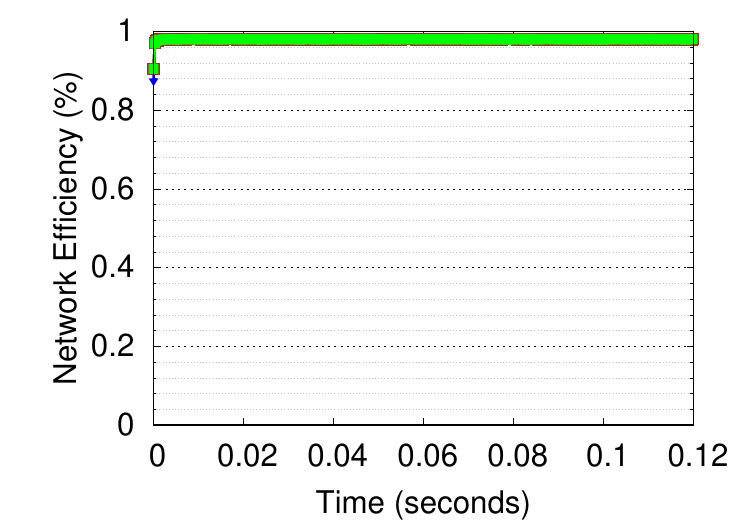}
\caption{RND - Vftree}
\label{fig_RLFT_RD_vftree3_T04_100}
\end{subfigure}
\begin{subfigure}{0.245\textwidth}
 \centering 
\includegraphics[width=1\textwidth]
{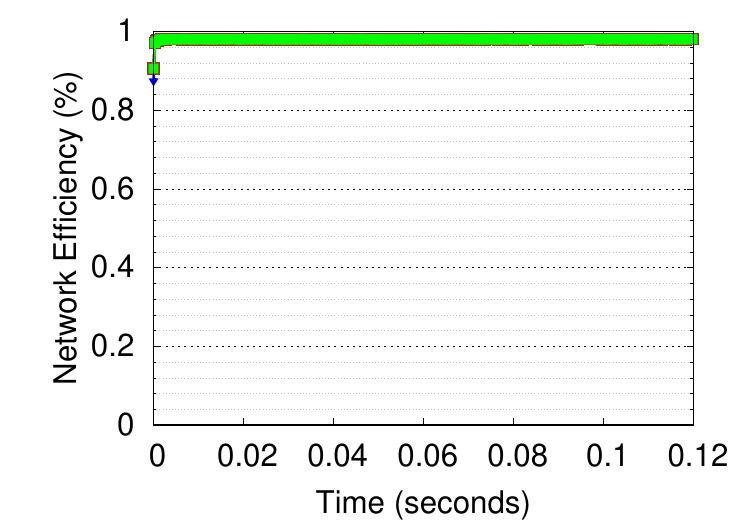}
\caption{RND - Flow2SL}
\label{fig_RLFT_RD_flow2sl3_T04_100}
\end{subfigure}

\begin{subfigure}{0.245\textwidth}
 \centering 
\includegraphics[width=1\textwidth]
{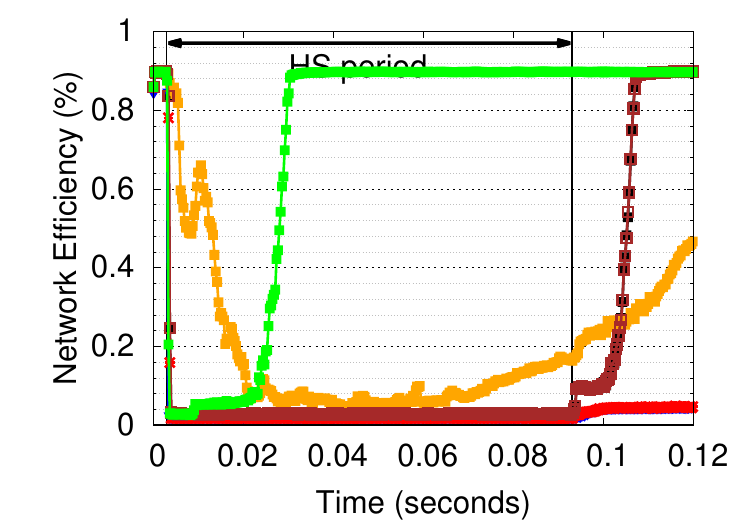}
\caption{H10- 1Q}
\label{fig_RLFT_HS10_1q_T04_100_3456}
\end{subfigure}
\begin{subfigure}{0.245\textwidth}
 \centering 
\includegraphics[width=1\textwidth]
{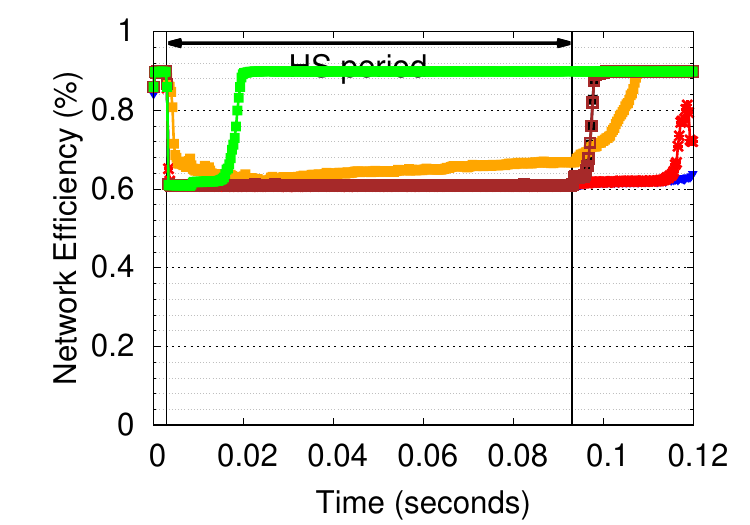}
\caption{H10- DBBM}
\label{fig_RLFT_HS10_dbbm3_T04_100_3456}
\end{subfigure}
\begin{subfigure}{0.245\textwidth}
 \centering 
\includegraphics[width=1\textwidth]
{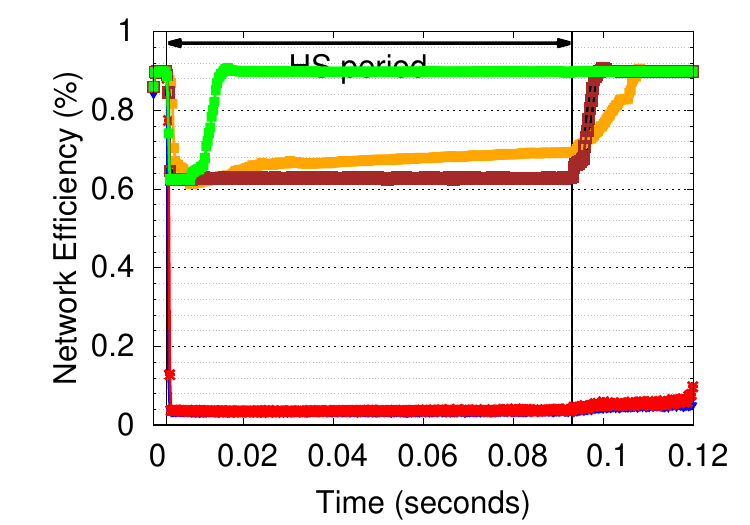}
\caption{H10- Vftree}
\label{fig_RLFT_HS10_vftree3_T04_100_3456}
\end{subfigure}
\begin{subfigure}{0.245\textwidth}
 \centering 
\includegraphics[width=1\textwidth]
{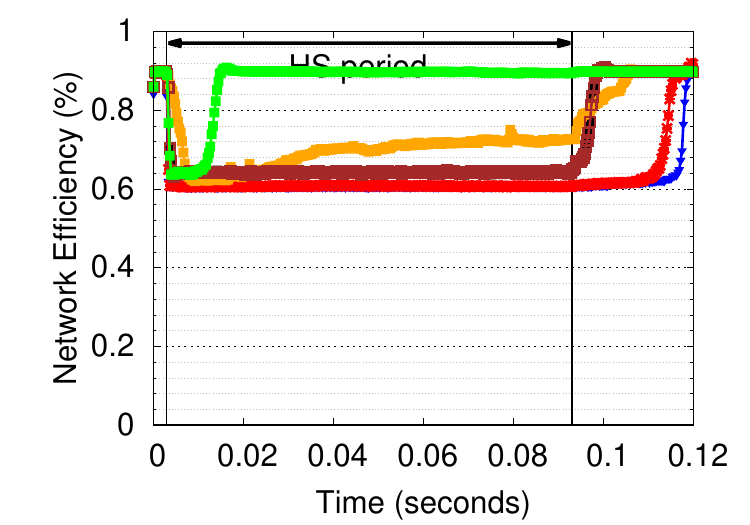}
\caption{H10- Flow2SL}
\label{fig_RLFT_HS10_flow2sl3_T04_100_3456}
\end{subfigure}

\begin{subfigure}{0.245\textwidth}
 \centering 
\includegraphics[width=1\textwidth]
{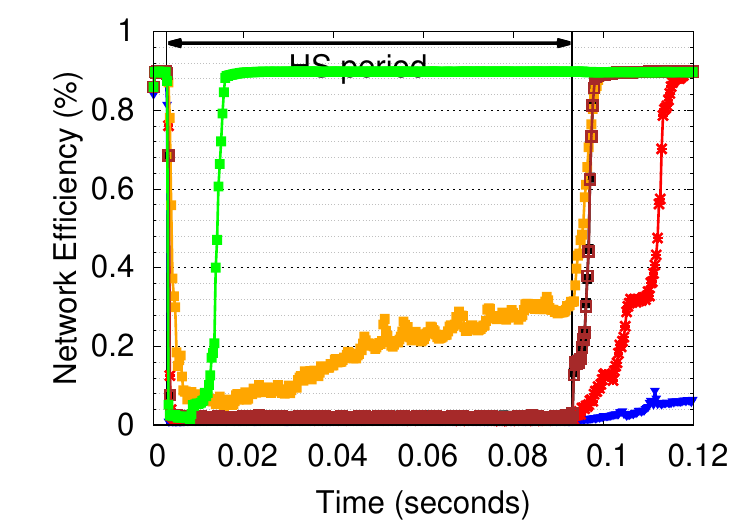}
\caption{H10-4 - 1Q}
\label{fig_RLFT_HS10-4_1q_T04_100_3456}
\end{subfigure}
\begin{subfigure}{0.245\textwidth}
 \centering 
\includegraphics[width=1\textwidth]
{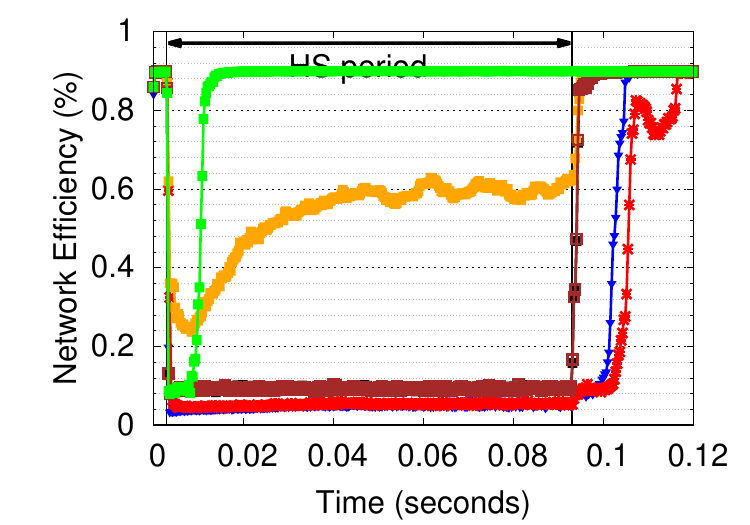}
\caption{H10-4 - DBBM}
\label{fig_RLFT_HS10-4_dbbm3_T04_100_3456}
\end{subfigure}
\begin{subfigure}{0.245\textwidth}
 \centering 
\includegraphics[width=1\textwidth]
{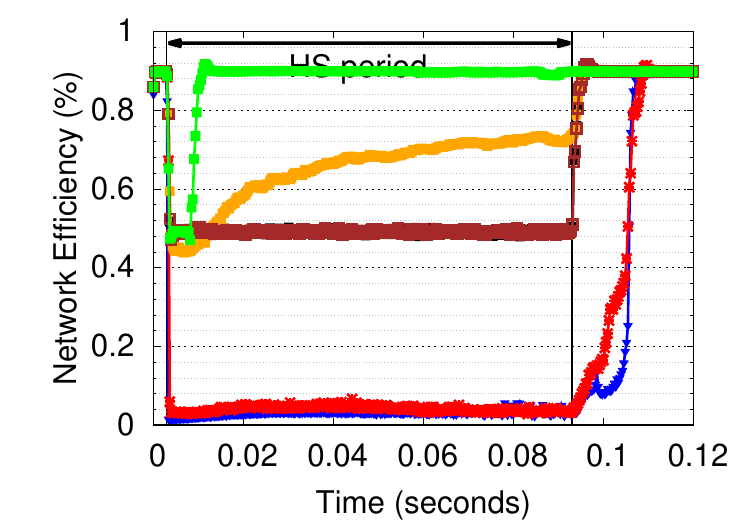}
\caption{H10-4 - Vftree}
\label{fig_RLFT_HS10-4_vftree3_T04_100_3456}
\end{subfigure}
\begin{subfigure}{0.245\textwidth}
 \centering 
\includegraphics[width=1\textwidth]
{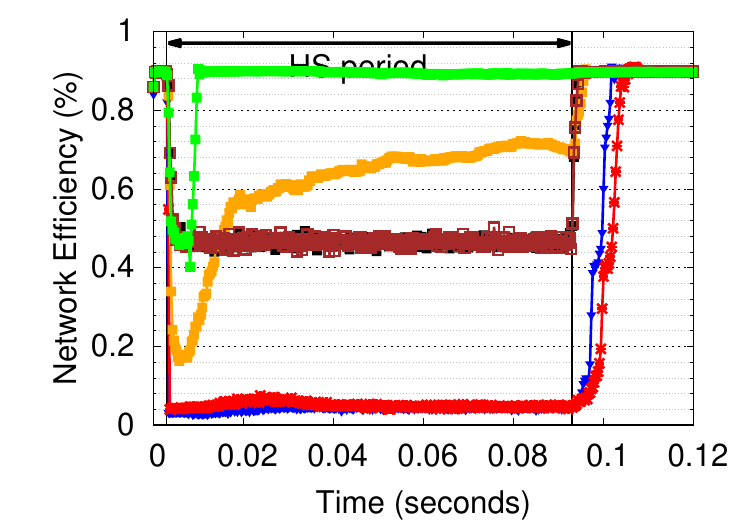}
\caption{H10-4 - Flow2SL}
\label{fig_RLFT_HS10-4_flow2sl3_T04_100_3456}
\end{subfigure}

\begin{subfigure}{0.245\textwidth}
 \centering 
\includegraphics[width=1\textwidth]
{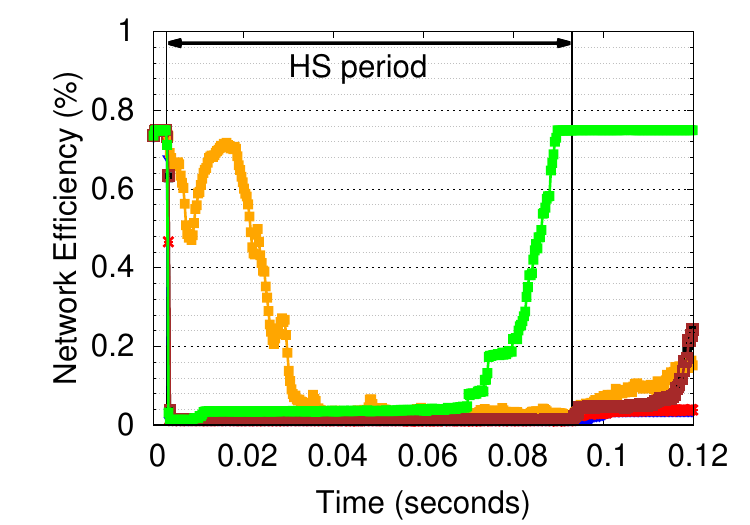}
\caption{H25  - 1Q}
\label{fig_RLFT_HS25_1q_T04_100_3456}
\end{subfigure}
\begin{subfigure}{0.245\textwidth}
 \centering 
\includegraphics[width=1\textwidth]
{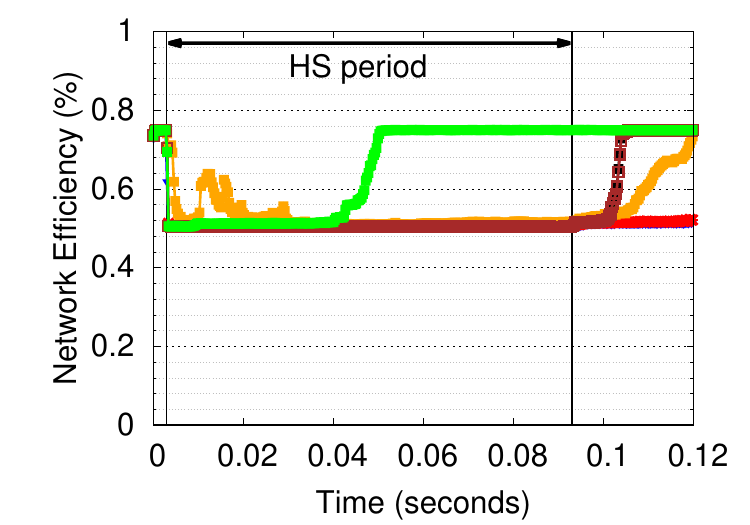}
\caption{H25  - DBBM}
\label{fig_RLFT_HS25_dbbm3_T04_100_3456}
\end{subfigure}
\begin{subfigure}{0.245\textwidth}
 \centering 
\includegraphics[width=1\textwidth]
{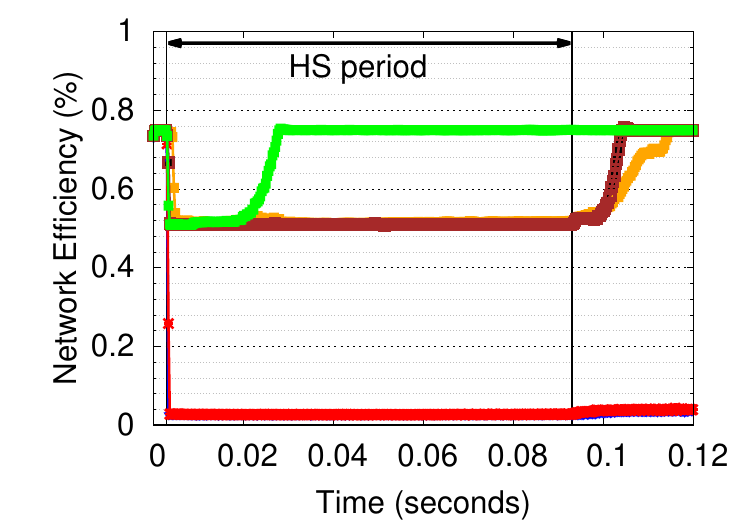}
\caption{H25  - Vftree}
\label{fig_RLFT_HS25_vftree3_T04_100_3456}
\end{subfigure}
\begin{subfigure}{0.245\textwidth}
 \centering 
\includegraphics[width=1\textwidth]
{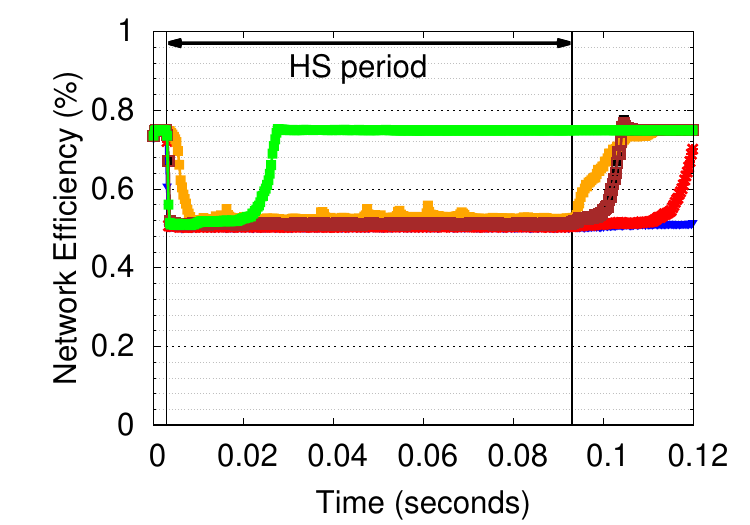}
\caption{H25  - Flow2SL}
\label{fig_RLFT_HS25_flow2sl3_T04_100_3456}
\end{subfigure}

\begin{subfigure}{0.245\textwidth}
 \centering 
\includegraphics[width=1\textwidth]
{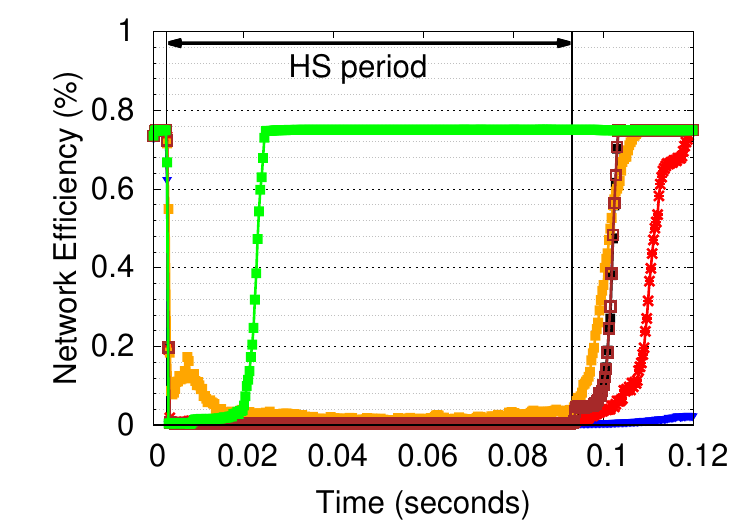}
\caption{H25-4 - 1Q}
\label{fig_RLFT_HS25-4_1q_T04_100_3456}
\end{subfigure}
\begin{subfigure}{0.245\textwidth}
 \centering 
\includegraphics[width=1\textwidth]
{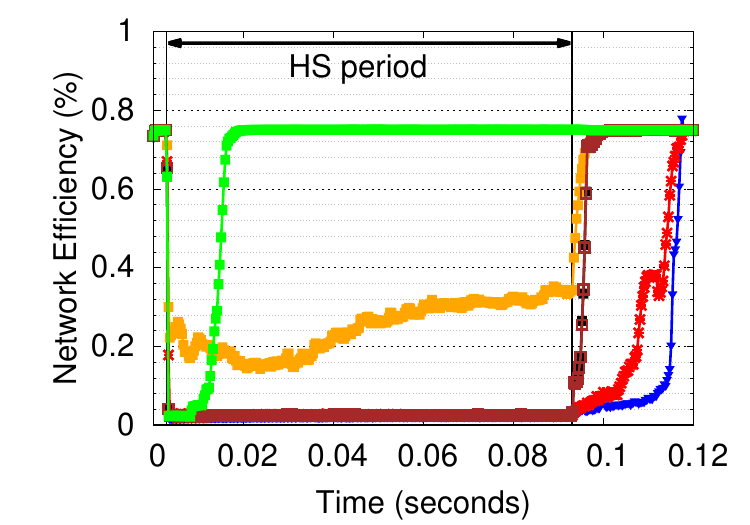}
\caption{H25-4 - DBBM}
\label{fig_RLFT_HS25-4_dbbm3_T04_100_3456}
\end{subfigure}
\begin{subfigure}{0.245\textwidth}
 \centering 
\includegraphics[width=1\textwidth]
{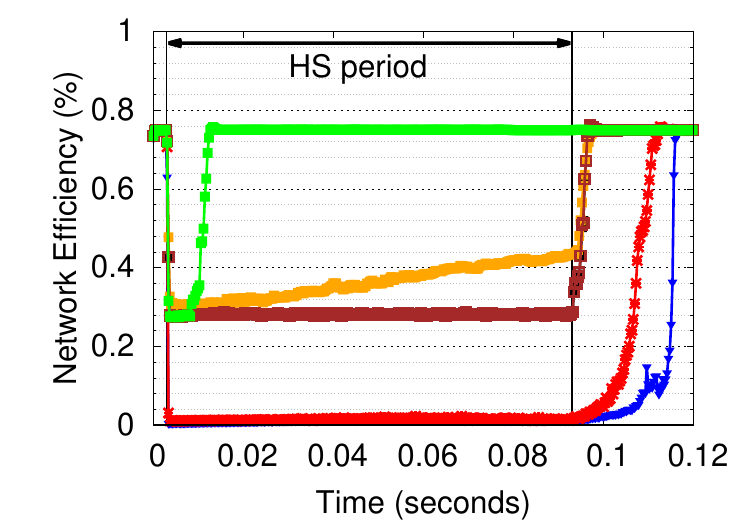}
\caption{H25-4 - Vftree}
\label{fig_RLFT_HS25-4_vftree3_T04_100_3456}
\end{subfigure}
\begin{subfigure}{0.245\textwidth}
 \centering 
\includegraphics[width=1\textwidth]
{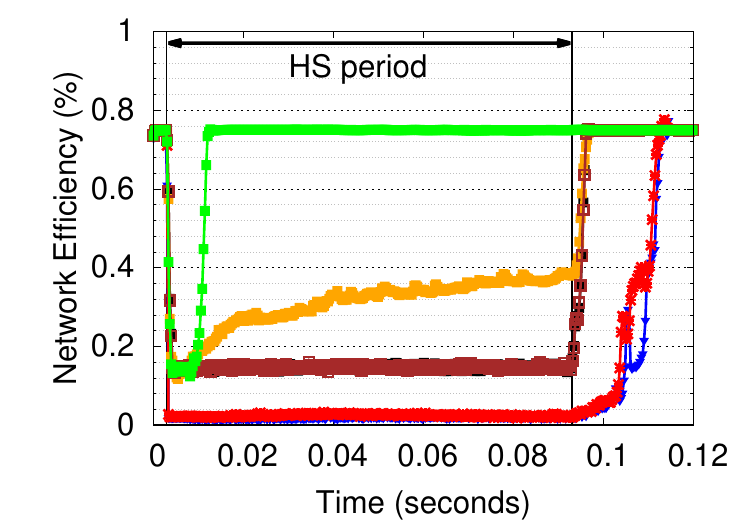}
\caption{H25-4 - Flow2SL}
\label{fig_RLFT_HS25-4_flow2sl3_T04_100_3456}
\end{subfigure}

\caption{Network Efficiency versus Time for the modeled techniques in a $3456$-node RLFT network (see configuration \#2 in Table \ref{tab:networks}), when traffic patterns described in Section \ref{sec:evaluation:trafic-model} are generated.} 
\label{fig:RLFT_3456_T04_100}
\end{figure*}

\figurename~\ref{fig:RLFT_3456_T04_100} shows the experiment results for the modeled techniques when we simulate a larger network configuration, where the number of interconnected endnodes increases up to $3456$ (see network configuration \#2 in Table \ref{tab:networks}), while the percentage of nodes contributing to congestion remains constant, i.e., 10\% and 25\% for the \emph{HS10} and \emph{HS25} traffic patterns, respectively. Note that, the link speed is also $100$~Gbps, like in the previous scenario.
Thus, we increase the overall number of congesting packets being injected in the network, compared to network configuration \#1. Hence, the network points that enter into congestion state is significantly larger in network configuration \#2 compared to \#1.
Therefore, as we have large congestion trees, the congestion effects will be very dramatic when adaptive routing ends up spreading the congestion trees and the congestion control technique is not able to react quickly. This configuration has been evaluated to demonstrate the scalability of \emph{ARN+AFI}.

In this context (see \figurename~\ref{fig:RLFT_3456_T04_100}), when the \emph{HS10} and \emph{HS25} traffic patterns are generated (i.e., only a single congestion tree is generated), we observe severe efficiency droppings, especially when no SQSs are implemented (see \figurename~\ref{fig_RLFT_HS10_1q_T04_100_3456} and ~\ref{fig_RLFT_HS25_1q_T04_100_3456}). In this case, \emph{ARN+AFI} is the only one that achieves to react against congestion, since it isolates congesting packets in the AFC space in the switch buffers, and the adaptive routing (i.e., the ARN mechanism) do not spread the congestion tree. However, note that \emph{ARN+AFI} takes longer to isolate the congested flows, compared to network configuration \#1, since the network size is also higher.
If we look at the results for \emph{HS10-4} and \emph{HS25-4} (see \figurename~\ref{fig_RLFT_HS10-4_1q_T04_100_3456} and \ref{fig_RLFT_HS25-4_1q_T04_100_3456}), we observe that the \emph{ARN+AFI} reacts much faster, despite having a number of branches similar to a single incast.
Note that the reason for that is the higher number of endnodes consuming congesting traffic ($4$ in this case), so that \emph{ARN+AFI} takes less time to drain the congestion trees.
However, note that in this case it benefits \emph{ARN+AFI} to be combined with a SQSs, so that it deals with HoL blocking during the reaction time to isolate congesting traffic.

When we use DBBM (see in Figures \ref{fig_RLFT_HS10_dbbm3_T04_100_3456} and \ref{fig_RLFT_HS25_dbbm3_T04_100_3456}), we notice that \emph{ARN+AFI} requires a reaction time higher than the other SQSs configurations for cleaning the VC clogged by congestion.
The DBBM policy for mapping packets to VCs can be very unfortunate in Fat-Tree networks, as we have observed before \cite{RocherJPDC21}.
The main reason is that the static criterion to assign packets to VCs does not consider neither the topology nor the routing algorithm.
By contrast, Vftree and Flow2SL are topology- and routing-aware. They were designed to leverage all available VCs and the traffic balancing provided by the routing algorithm. This behavior makes them more efficient than DBBM. 
The rest of the sub-figures in \figurename~\ref{fig:RLFT_3456_T04_100} show that the reaction time of \emph{ARN+AFI} combined to VFtree or Flow2SL is shorter than that of DBBM, and the performance dropping is lower in many cases for the former ones, compared to the latter.

\begin{figure*}[bth]
 \centering 
\begin{subfigure}{.7\textwidth}
 \centering 
\includegraphics[width=1\columnwidth]
{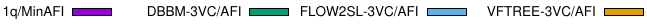}
\label{leyendaTrazas}
\end{subfigure}
\\
\vspace{-.4cm}
\begin{subfigure}{0.89\textwidth}
 \centering 
\includegraphics[width=1\columnwidth]
{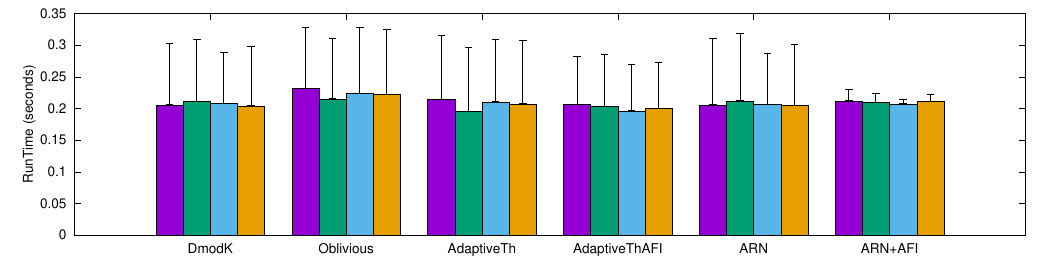}
\caption{PTRANS}
\label{fig_ptrans}
\end{subfigure}

\begin{subfigure}{0.89\textwidth}
 \centering 
\includegraphics[width=1\columnwidth]
{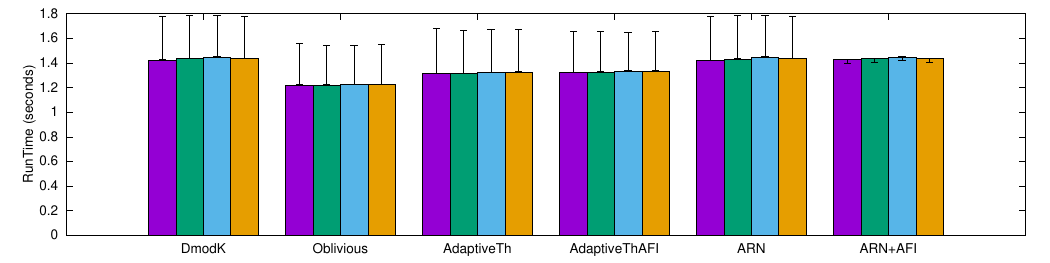}
\caption{Inception-V3}
\label{fig_inception}
\end{subfigure}
\caption{Execution time results for the trace-based traffic in a $432$-node RLFT network (see configuration \#1 in Table 2). Note that using AFI means to use an additional VC. Moreover, error bars show the overhead time when running in the presence of a congestion tree (see Section \ref{sec:evaluation:trafic-model})}
\label{fig:432_traces}
\end{figure*}

\subsection{Experiment results with trace-based traffic}
\label{sec:evaluation:traces}

\figurename~\ref{fig_ptrans} shows the execution time (in seconds) for the modeled queuing schemes combined with the routing algorithms, using the PTRANS test as a workload. 
Note that error bars show the overhead time of this application when running in the presence of a congestion tree (see Section \ref{sec:evaluation:trafic-model}), which lasts for $100$ ms.
As can be seen, this congestion tree delays the execution of PTRANS application for all the congestion control techniques but \emph{ARN+AFI}.
It is important to mention that PTRANS generates a uniform-like traffic pattern with small bursts of congestion. Indeed, this traffic pattern generates low-order HoL blocking but does not produce a significant high-order HoL blocking.
The \emph{Oblivious} and \emph{AdaptiveTh} routing algorithms spread the flows in the upward phase in the fat-tree topology, increasing the probability if high-order HoL blocking.
Indeed, the \emph{Oblivious} routing achieves the worst results (lower is better) regardless of the buffer organization, when no congestion is present in the network.
The rest of the techniques are able to execute the PTRANS test in about $0.2$ seconds, without significant variations.
However, when we a congestion tree is present in the network, we observe that the runtime of all the techniques significantly increases (see the error bars), except for \emph{ARN+AFI}, which obtains the best results, as it is able to reduce congestion spreading.

\figurename ~\ref{fig_inception} shows the execution time (in seconds) when using the Inception-V3 model with the same configuration as mentioned above for PTRANS. In this case, \emph{Oblivious}, \emph{AdaptiveTh} and \emph{AdaptiveTHAFI} improve \emph{$D$-mod-$K$}, since the congestion in this traffic patter running alone is light, and so high-order HoL blocking presence is small.
However, when the congestion tree lasting for $400$ms is generated (see Section \ref{sec:evaluation:trafic-model}) combined with the Inception-V3 application, the execution time overhead is close to the congestion tree time for all the routing configurations, except for \emph{ARN+AFI}, which is able to precisely identify the congesting flows and isolate them in the AFC to reduce the congestion spreading.
Note also that the overhead introduced by congestion trees in both applications (i.e., PTRANS and Inception-v3) is close to zero, which means that \emph{ARN+AFI} is an efficient technique to reduce the congestion interference in the application execution time.

\section{Conclusion}
\label{sec:conclusions}

In this paper, we have presented ARN$+$AFI, a new technique to mitigate the impact of congestion spreading throughout the network when adaptive routing is used.
ARN$+$AFI detects congestion roots, and adapts the traffic flows in a smart way using the ARN mechanism, available in InfiniBand networks (or a similar one as long as it exchanges congestion notifications among switches).
ARN$+$AFI isolates the traffic flows being routed through alternative routes into a special virtual channel (VC), called adapted flow channel (AFC), while other flows remain separated in regular VCs.
ARN$+$AFI can be combined with static queuing schemes (SQSs) to mitigate the Head-of-Line (HoL) blocking effect, derived from congestion situations.
We have conducted an extensive set of simulation experiments to evaluate our proposal in Fat-Tree topologies, modeling realistic traffic patterns (both synthetic and trace-based).
The analysis of the obtained results shows that ARN$+$AFI avoids non-adapted traffic flows suffering from HoL blocking due to the congestion spreading.
Furthermore, the results also show that ARN$+$AFI combined with SQSs achieves a shorter reaction time, especially under heavily congested scenarios.
In general, the congestion impact is eliminated in the evaluated scenarios but introduces a delay depending on the size of the congestion tree.
For future work, we plan to evaluate the proposed approach under congestion scenarios other than incast, such as in-network congestion, where congestion points appear in network internal points.

\section*{Data Availability}
The datasets generated during and/or analysed during the current study are available from the corresponding author on reasonable request.

\section*{Conflict of interest}
All authors declare that they have no conflicts of interest.

\section*{Acknowledgments}
This work is part of the R\&D Project Grant PID2019-109001RA-I00, funded by MCIN/AEI/10.13039/501100011033.
 Moreover, this work has also been jointly supported by Junta de Comunidades de Castilla-La Mancha under the project SBPLY/17/180501/000498.

\bibliography{sn-bibliography}


\begin{thebibliography}{48}
\ifx \bisbn   \undefined \def \bisbn  #1{ISBN #1}\fi
\ifx \binits  \undefined \def \binits#1{#1}\fi
\ifx \bauthor  \undefined \def \bauthor#1{#1}\fi
\ifx \batitle  \undefined \def \batitle#1{#1}\fi
\ifx \bjtitle  \undefined \def \bjtitle#1{#1}\fi
\ifx \bvolume  \undefined \def \bvolume#1{\textbf{#1}}\fi
\ifx \byear  \undefined \def \byear#1{#1}\fi
\ifx \bissue  \undefined \def \bissue#1{#1}\fi
\ifx \bfpage  \undefined \def \bfpage#1{#1}\fi
\ifx \blpage  \undefined \def \blpage #1{#1}\fi
\ifx \burl  \undefined \def \burl#1{\textsf{#1}}\fi
\ifx \doiurl  \undefined \def \doiurl#1{\url{https://doi.org/#1}}\fi
\ifx \betal  \undefined \def \betal{\textit{et al.}}\fi
\ifx \binstitute  \undefined \def \binstitute#1{#1}\fi
\ifx \binstitutionaled  \undefined \def \binstitutionaled#1{#1}\fi
\ifx \bctitle  \undefined \def \bctitle#1{#1}\fi
\ifx \beditor  \undefined \def \beditor#1{#1}\fi
\ifx \bpublisher  \undefined \def \bpublisher#1{#1}\fi
\ifx \bbtitle  \undefined \def \bbtitle#1{#1}\fi
\ifx \bedition  \undefined \def \bedition#1{#1}\fi
\ifx \bseriesno  \undefined \def \bseriesno#1{#1}\fi
\ifx \blocation  \undefined \def \blocation#1{#1}\fi
\ifx \bsertitle  \undefined \def \bsertitle#1{#1}\fi
\ifx \bsnm \undefined \def \bsnm#1{#1}\fi
\ifx \bsuffix \undefined \def \bsuffix#1{#1}\fi
\ifx \bparticle \undefined \def \bparticle#1{#1}\fi
\ifx \barticle \undefined \def \barticle#1{#1}\fi
\bibcommenthead
\ifx \bconfdate \undefined \def \bconfdate #1{#1}\fi
\ifx \botherref \undefined \def \botherref #1{#1}\fi
\ifx \url \undefined \def \url#1{\textsf{#1}}\fi
\ifx \bchapter \undefined \def \bchapter#1{#1}\fi
\ifx \bbook \undefined \def \bbook#1{#1}\fi
\ifx \bcomment \undefined \def \bcomment#1{#1}\fi
\ifx \oauthor \undefined \def \oauthor#1{#1}\fi
\ifx \citeauthoryear \undefined \def \citeauthoryear#1{#1}\fi
\ifx \endbibitem  \undefined \def \endbibitem {}\fi
\ifx \bconflocation  \undefined \def \bconflocation#1{#1}\fi
\ifx \arxivurl  \undefined \def \arxivurl#1{\textsf{#1}}\fi
\csname PreBibitemsHook\endcsname

\bibitem{FatTree}
\begin{barticle}
\bauthor{\bsnm{Leiserson}, \binits{C.E.}}:
\batitle{Fat-trees: Universal networks for hardware-efficient supercomputing}.
\bjtitle{IEEE Transactions on Computers}
\bvolume{C-34}(\bissue{10}),
\bfpage{892}--\blpage{901}
(\byear{1985}).
\doiurl{10.1109/TC.1985.6312192}
\end{barticle}
\endbibitem

\bibitem{Gomez07ipdps}
\begin{bchapter}
\bauthor{\bsnm{Requena}, \binits{C.G.}},
\bauthor{\bsnm{Villam{\'o}n}, \binits{F.G.}},
\bauthor{\bsnm{G{\'o}mez}, \binits{M.E.}},
\bauthor{\bsnm{L{\'o}pez}, \binits{P.}},
\bauthor{\bsnm{Duato}, \binits{J.}}:
\bctitle{{Deterministic versus Adaptive Routing in Fat-Trees}}.
In: \bbtitle{{21th International Parallel and Distributed Processing Symposium
  {(IPDPS) 2007}, Proceedings, 26-30 March 2007, Long Beach, California,
  {USA}}},
pp. \bfpage{1}--\blpage{8}.
\bpublisher{{IEEE}}, \blocation{???}
(\byear{2007}).
\doiurl{10.1109/IPDPS.2007.370482}
\end{bchapter}
\endbibitem

\bibitem{Zahavi2010}
\begin{barticle}
\bauthor{\bsnm{Zahavi}, \binits{E.}},
\bauthor{\bsnm{Johnson}, \binits{G.}},
\bauthor{\bsnm{Kerbyson}, \binits{D.J.}},
\bauthor{\bsnm{Lang}, \binits{M.}}:
\batitle{{Optimized {InfiniBand}$^{\mbox{TM}}$ fat-tree routing for shift
  all-to-all communication patterns}}.
\bjtitle{Journal of CCPE}
\bvolume{22}(\bissue{2}),
\bfpage{217}--\blpage{231}
(\byear{2010})
\end{barticle}
\endbibitem

\bibitem{ObliviousFT}
\begin{bchapter}
\bauthor{\bsnm{Rodriguez}, \binits{G.}},
\bauthor{\bsnm{Minkenberg}, \binits{C.}},
\bauthor{\bsnm{Beivide}, \binits{R.}},
\bauthor{\bsnm{Luijten}, \binits{R.P.}},
\bauthor{\bsnm{Labarta}, \binits{J.}},
\bauthor{\bsnm{Valero}, \binits{M.}}:
\bctitle{Oblivious routing schemes in extended generalized fat tree networks}.
In: \bbtitle{2009 IEEE International Conference on Cluster Computing and
  Workshops},
pp. \bfpage{1}--\blpage{8}
(\byear{2009}).
\doiurl{10.1109/CLUSTR.2009.5289145}
\end{bchapter}
\endbibitem

\bibitem{DBLP:journals/jsac/ZahaviKK14}
\begin{barticle}
\bauthor{\bsnm{Zahavi}, \binits{E.}},
\bauthor{\bsnm{Keslassy}, \binits{I.}},
\bauthor{\bsnm{Kolodny}, \binits{A.}}:
\batitle{Distributed adaptive routing convergence to non-blocking {DCN} routing
  assignments}.
\bjtitle{{IEEE} Journal on Selected Areas in Communications}
\bvolume{32}(\bissue{1}),
\bfpage{88}--\blpage{101}
(\byear{2014}).
\doiurl{10.1109/JSAC.2014.140109}
\end{barticle}
\endbibitem

\bibitem{DBLP:conf/hoti/GeoffrayH08}
\begin{bchapter}
\bauthor{\bsnm{Geoffray}, \binits{P.}},
\bauthor{\bsnm{Hoefler}, \binits{T.}}:
\bctitle{Adaptive routing strategies for modern high performance networks}.
In: \bbtitle{16th Annual {IEEE} Symposium on High Performance Interconnects
  ({HOTI} 2008), 26-28 August 2008, Stanford, CA, {USA}},
pp. \bfpage{165}--\blpage{172}.
\bpublisher{{IEEE} Computer Society}, \blocation{???}
(\byear{2008}).
\doiurl{10.1109/HOTI.2008.21}.
\burl{http://dx.doi.org/10.1109/HOTI.2008.21}
\end{bchapter}
\endbibitem

\bibitem{DBLP:conf/sc/KimDA06}
\begin{bchapter}
\bauthor{\bsnm{Kim}, \binits{J.}},
\bauthor{\bsnm{Dally}, \binits{W.J.}},
\bauthor{\bsnm{Abts}, \binits{D.}}:
\bctitle{Interconnect routing and scheduling - adaptive routing in high-radix
  clos network}.
In: \bbtitle{Proceedings of the {ACM/IEEE} {SC2006} Conference on High
  Performance Networking and Computing, November 11-17, 2006, Tampa, FL,
  {USA}},
p. \bfpage{92}.
\bpublisher{{ACM} Press}, \blocation{???}
(\byear{2006}).
\doiurl{10.1145/1188455.1188552}.
\burl{http://doi.acm.org/10.1145/1188455.1188552}
\end{bchapter}
\endbibitem

\bibitem{adaptiveRCA}
\begin{bchapter}
\bauthor{\bsnm{Gratz}, \binits{P.}},
\bauthor{\bsnm{Grot}, \binits{B.}},
\bauthor{\bsnm{Keckler}, \binits{S.W.}}:
\bctitle{Regional congestion awareness for load balance in networks-on-chip}.
In: \bbtitle{2008 IEEE 14th International Symposium on High Performance
  Computer Architecture},
pp. \bfpage{203}--\blpage{214}
(\byear{2008}).
\doiurl{10.1109/HPCA.2008.4658640}
\end{bchapter}
\endbibitem

\bibitem{adaptiveDBAR}
\begin{bchapter}
\bauthor{\bsnm{Ma}, \binits{S.}},
\bauthor{\bsnm{Jerger}, \binits{N.E.}},
\bauthor{\bsnm{Wang}, \binits{Z.}}:
\bctitle{Dbar: An efficient routing algorithm to support multiple concurrent
  applications in networks-on-chip}.
In: \bbtitle{2011 38th Annual International Symposium on Computer Architecture
  (ISCA)},
pp. \bfpage{413}--\blpage{424}
(\byear{2011})
\end{bchapter}
\endbibitem

\bibitem{ecmp}
\begin{barticle}
\bauthor{\bsnm{Hopps}, \binits{C.E.}}:
\batitle{Analysis of an equal-cost multi-path algorithm}.
\bjtitle{{RFC}}
\bvolume{2992},
\bfpage{1}--\blpage{8}
(\byear{2000}).
\doiurl{10.17487/RFC2992}
\end{barticle}
\endbibitem

\bibitem{presto}
\begin{bchapter}
\bauthor{\bsnm{He}, \binits{K.}},
\bauthor{\bsnm{Rozner}, \binits{E.}},
\bauthor{\bsnm{Agarwal}, \binits{K.}},
\bauthor{\bsnm{Felter}, \binits{W.}},
\bauthor{\bsnm{Carter}, \binits{J.B.}},
\bauthor{\bsnm{Akella}, \binits{A.}}:
\bctitle{Presto: Edge-based load balancing for fast datacenter networks}.
In: \beditor{\bsnm{Uhlig}, \binits{S.}},
\beditor{\bsnm{Maennel}, \binits{O.}},
\beditor{\bsnm{Karp}, \binits{B.}},
\beditor{\bsnm{Padhye}, \binits{J.}} (eds.)
\bbtitle{Proceedings of the 2015 {ACM} Conference on Special Interest Group on
  Data Communication, {SIGCOMM} 2015, London, United Kingdom, August 17-21,
  2015},
pp. \bfpage{465}--\blpage{478}.
\bpublisher{{ACM}}, \blocation{???}
(\byear{2015}).
\doiurl{10.1145/2785956.2787507}.
\burl{https://doi.org/10.1145/2785956.2787507}
\end{bchapter}
\endbibitem

\bibitem{drill}
\begin{bchapter}
\bauthor{\bsnm{Ghorbani}, \binits{S.}},
\bauthor{\bsnm{Yang}, \binits{Z.}},
\bauthor{\bsnm{Godfrey}, \binits{P.B.}},
\bauthor{\bsnm{Ganjali}, \binits{Y.}},
\bauthor{\bsnm{Firoozshahian}, \binits{A.}}:
\bctitle{{DRILL:} micro load balancing for low-latency data center networks}.
In: \bbtitle{Proceedings of the Conference of the {ACM} Special Interest Group
  on Data Communication, {SIGCOMM} 2017, Los Angeles, CA, USA, August 21-25,
  2017},
pp. \bfpage{225}--\blpage{238}.
\bpublisher{{ACM}}, \blocation{???}
(\byear{2017}).
\doiurl{10.1145/3098822.3098839}.
\burl{https://doi.org/10.1145/3098822.3098839}
\end{bchapter}
\endbibitem

\bibitem{power2}
\begin{barticle}
\bauthor{\bsnm{Wang}, \binits{S.}},
\bauthor{\bsnm{Luo}, \binits{J.}},
\bauthor{\bsnm{Wong}, \binits{W.S.}}:
\batitle{Improved power of two choices for fat-tree routing}.
\bjtitle{IEEE Transactions on Network and Service Management}
\bvolume{15}(\bissue{4}),
\bfpage{1706}--\blpage{1719}
(\byear{2018}).
\doiurl{10.1109/TNSM.2018.2865543}
\end{barticle}
\endbibitem

\bibitem{Besta14sc}
\begin{bchapter}
\bauthor{\bsnm{Besta}, \binits{M.}},
\bauthor{\bsnm{Hoefler}, \binits{T.}}:
\bctitle{{Slim Fly: {A} Cost Effective Low-Diameter Network Topology}}.
In: \beditor{\bsnm{Damkroger}, \binits{T.}},
\beditor{\bsnm{Dongarra}, \binits{J.}} (eds.)
\bbtitle{{International Conference for High Performance Computing, Networking,
  Storage and Analysis, {SC} 2014, New Orleans, LA, USA, November 16-21,
  2014}},
pp. \bfpage{348}--\blpage{359}.
\bpublisher{{IEEE}}, \blocation{???}
(\byear{2014}).
\doiurl{10.1109/SC.2014.34}
\end{bchapter}
\endbibitem

\bibitem{Jellyfish}
\begin{bchapter}
\bauthor{\bsnm{Singla}, \binits{A.}},
\bauthor{\bsnm{Hong}, \binits{C.}},
\bauthor{\bsnm{Popa}, \binits{L.}},
\bauthor{\bsnm{Godfrey}, \binits{P.B.}}:
\bctitle{Jellyfish: Networking data centers randomly}.
In: \beditor{\bsnm{Gribble}, \binits{S.D.}},
\beditor{\bsnm{Katabi}, \binits{D.}} (eds.)
\bbtitle{Proceedings of the 9th {USENIX} Symposium on Networked Systems Design
  and Implementation, {NSDI} 2012, San Jose, CA, USA, April 25-27, 2012},
pp. \bfpage{225}--\blpage{238}.
\bpublisher{{USENIX} Association}, \blocation{???}
(\byear{2012}).
\burl{https://www.usenix.org/conference/nsdi12/technical-sessions/presentation/singla}
\end{bchapter}
\endbibitem

\bibitem{Kim08isca}
\begin{bchapter}
\bauthor{\bsnm{Kim}, \binits{J.}},
\bauthor{\bsnm{Dally}, \binits{W.J.}},
\bauthor{\bsnm{Scott}, \binits{S.}},
\bauthor{\bsnm{Abts}, \binits{D.}}:
\bctitle{{Technology-Driven, Highly-Scalable Dragonfly Topology}}.
In: \bbtitle{{35th International Symposium on Computer Architecture {(ISCA)
  2008}, June 21-25, 2008, Beijing, China}},
pp. \bfpage{77}--\blpage{88}.
\bpublisher{{IEEE} Computer Society}, \blocation{???}
(\byear{2008}).
\doiurl{10.1109/ISCA.2008.19}
\end{bchapter}
\endbibitem

\bibitem{Valiant82siamcomp}
\begin{barticle}
\bauthor{\bsnm{Valiant}, \binits{L.G.}}:
\batitle{{A Scheme for Fast Parallel Communication}}.
\bjtitle{{SIAM} J. Comput.}
\bvolume{11}(\bissue{2}),
\bfpage{350}--\blpage{361}
(\byear{1982}).
\doiurl{10.1137/0211027}
\end{barticle}
\endbibitem

\bibitem{Jiang09}
\begin{bchapter}
\bauthor{\bsnm{Jiang}, \binits{N.}},
\bauthor{\bsnm{Kim}, \binits{J.}},
\bauthor{\bsnm{Dally}, \binits{W.J.}}:
\bctitle{{Indirect adaptive routing on large scale interconnection networks}}.
In: \bbtitle{{36th International Symposium on Computer Architecture ({ISCA}
  2009), June 20-24, 2009, Austin, TX, {USA}}},
pp. \bfpage{220}--\blpage{231}
(\byear{2009}).
\doiurl{10.1145/1555754.1555783}
\end{bchapter}
\endbibitem

\bibitem{Newaz21}
\begin{bchapter}
\bauthor{\bsnm{Newaz}, \binits{M.N.}},
\bauthor{\bsnm{Mollah}, \binits{M.A.}},
\bauthor{\bsnm{Faizian}, \binits{P.}},
\bauthor{\bsnm{Tong}, \binits{Z.}}:
\bctitle{Improving adaptive routing performance on large scale megafly
  topology}.
In: \bbtitle{The 21st IEEE/ACM International Symposium on Cluster, Cloud and
  Internet Computing, {CCGrid}, May 10-13, 2021, Melbourne, Victoria,
  Australia},
p. \bfpage{1}.
\bpublisher{{IEEE/ACM}}, \blocation{???}
(\byear{2021})
\end{bchapter}
\endbibitem

\bibitem{multiPathTable}
\begin{barticle}
\bauthor{\bsnm{Besta}, \binits{M.}},
\bauthor{\bsnm{Domke}, \binits{J.}},
\bauthor{\bsnm{Schneider}, \binits{M.}},
\bauthor{\bsnm{Konieczny}, \binits{M.}},
\bauthor{\bsnm{Girolamo}, \binits{S.D.}},
\bauthor{\bsnm{Schneider}, \binits{T.}},
\bauthor{\bsnm{Singla}, \binits{A.}},
\bauthor{\bsnm{Hoefler}, \binits{T.}}:
\batitle{High-performance routing with multipathing and path diversity in
  ethernet and {HPC} networks}.
\bjtitle{{IEEE} Trans. Parallel Distributed Syst.}
\bvolume{32}(\bissue{4}),
\bfpage{943}--\blpage{959}
(\byear{2021}).
\doiurl{10.1109/TPDS.2020.3035761}
\end{barticle}
\endbibitem

\bibitem{Congdon21Nendica}
\begin{botherref}
\oauthor{\bsnm{Guo}, \binits{L.}},
\oauthor{\bsnm{Congdon}, \binits{P.}}:
Ieee 802 nendica report: Intelligent lossless data center networks.
IEEE SA Industry Connections--IEEE 802 Nendica Report: Intelligent Lossless
  Data Center Networks,
1--44
(2021)
\end{botherref}
\endbibitem

\bibitem{Garcia05hipeac}
\begin{bchapter}
\bauthor{\bsnm{Garc{\'{\i}}a}, \binits{P.J.}},
\bauthor{\bsnm{Flich}, \binits{J.}},
\bauthor{\bsnm{Duato}, \binits{J.}},
\bauthor{\bsnm{Johnson}, \binits{I.}},
\bauthor{\bsnm{Quiles}, \binits{F.J.}},
\bauthor{\bsnm{Naven}, \binits{F.}}:
\bctitle{Dynamic evolution of congestion trees: Analysis and impact on switch
  architecture}.
In: \bbtitle{High Performance Embedded Architectures and Compilers, First
  International Conference, HiPEAC 2005, Barcelona, Spain, November 17-18,
  2005, Proceedings},
pp. \bfpage{266}--\blpage{285}
(\byear{2005}).
\doiurl{10.1007/11587514_18}.
\burl{https://doi.org/10.1007/11587514\_18}
\end{bchapter}
\endbibitem

\bibitem{DBLP:journals/tcom/KarolHM87}
\begin{barticle}
\bauthor{\bsnm{Karol}, \binits{M.J.}},
\bauthor{\bsnm{Hluchyj}, \binits{M.G.}},
\bauthor{\bsnm{Morgan}, \binits{S.P.}}:
\batitle{Input versus output queueing on a space-division packet switch}.
\bjtitle{{IEEE} Trans. Communications}
\bvolume{35}(\bissue{12}),
\bfpage{1347}--\blpage{1356}
(\byear{1987}).
\doiurl{10.1109/TCOM.1987.1096719}
\end{barticle}
\endbibitem

\bibitem{bufferhogging}
\begin{bchapter}
\bauthor{\bsnm{Yoshigoe}, \binits{K.}}:
\bctitle{Threshold-based exhaustive round-robin for the cicq switch with
  virtual crosspoint queues}.
In: \bbtitle{2007 IEEE International Conference on Communications},
pp. \bfpage{6325}--\blpage{6329}
(\byear{2007}).
\doiurl{10.1109/ICC.2007.1047}
\end{bchapter}
\endbibitem

\bibitem{Jurczyk96phenomenonof}
\begin{botherref}
\oauthor{\bsnm{Jurczyk}, \binits{M.}},
\oauthor{\bsnm{Schwederski}, \binits{T.}}:
Phenomenon of Higher Order Head-of-Line Blocking in Multistage Interconnection
  Networks under Nonuniform Traffic Patterns
(1996)
\end{botherref}
\endbibitem

\bibitem{RocherJPDC21}
\begin{barticle}
\bauthor{\bsnm{Rocher{-}Gonzalez}, \binits{J.}},
\bauthor{\bsnm{Escudero{-}Sahuquillo}, \binits{J.}},
\bauthor{\bsnm{Garc{\'{\i}}a}, \binits{P.J.}},
\bauthor{\bsnm{Quiles}, \binits{F.J.}},
\bauthor{\bsnm{Mora}, \binits{G.}}:
\batitle{Towards an efficient combination of adaptive routing and queuing
  schemes in fat-tree topologies}.
\bjtitle{J. Parallel Distributed Comput.}
\bvolume{147},
\bfpage{46}--\blpage{63}
(\byear{2021}).
\doiurl{10.1016/j.jpdc.2020.07.009}
\end{barticle}
\endbibitem

\bibitem{rocherCCgrid}
\begin{bchapter}
\bauthor{\bsnm{Rocher{-}Gonzalez}, \binits{J.}},
\bauthor{\bsnm{Escudero{-}Sahuquillo}, \binits{J.}},
\bauthor{\bsnm{Garc{\'{\i}}a}, \binits{P.J.}},
\bauthor{\bsnm{Flor}, \binits{F.J.Q.}},
\bauthor{\bsnm{Mora}, \binits{G.}}:
\bctitle{Efficient congestion management for high-speed interconnects using
  adaptive routing}.
In: \bbtitle{19th {IEEE/ACM} International Symposium on Cluster, Cloud and Grid
  Computing, {CCGRID} 2019, Larnaca, Cyprus, May 14-17, 2019},
pp. \bfpage{221}--\blpage{230}.
\bpublisher{{IEEE}}, \blocation{???}
(\byear{2019}).
\doiurl{10.1109/CCGRID.2019.00036}.
\burl{https://doi.org/10.1109/CCGRID.2019.00036}
\end{bchapter}
\endbibitem

\bibitem{Nachiondo10pds}
\begin{barticle}
\bauthor{\bsnm{Nachiondo}, \binits{T.}},
\bauthor{\bsnm{Flich}, \binits{J.}},
\bauthor{\bsnm{Duato}, \binits{J.}}:
\batitle{{Buffer Management Strategies to Reduce HoL Blocking}}.
\bjtitle{Parallel and Distributed Systems, IEEE Transactions on}
\bvolume{21}(\bissue{6}),
\bfpage{739}--\blpage{753}
(\byear{2010}).
\doiurl{10.1109/TPDS.2009.63}
\end{barticle}
\endbibitem

\bibitem{Guay11ipdps}
\begin{bchapter}
\bauthor{\bsnm{Guay}, \binits{W.L.}},
\bauthor{\bsnm{Bogdanski}, \binits{B.}},
\bauthor{\bsnm{Reinemo}, \binits{S.}},
\bauthor{\bsnm{Lysne}, \binits{O.}},
\bauthor{\bsnm{Skeie}, \binits{T.}}:
\bctitle{{vFtree - {A} Fat-Tree Routing Algorithm Using Virtual Lanes to
  Alleviate Congestion}}.
In: \bbtitle{{25th {IEEE} International Symposium on Parallel and Distributed
  Processing, {IPDPS} 2011, Anchorage, Alaska, USA, 16-20 May, 2011 -
  Conference Proceedings}},
pp. \bfpage{197}--\blpage{208}
(\byear{2011}).
\doiurl{10.1109/IPDPS.2011.28}
\end{bchapter}
\endbibitem

\bibitem{Escudero14jpdc}
\begin{barticle}
\bauthor{\bsnm{Escudero{-}Sahuquillo}, \binits{J.}},
\bauthor{\bsnm{Garc{\'i}a}, \binits{P.J.}},
\bauthor{\bsnm{Quiles}, \binits{F.J.}},
\bauthor{\bsnm{Reinemo}, \binits{S.}},
\bauthor{\bsnm{Skeie}, \binits{T.}},
\bauthor{\bsnm{Lysne}, \binits{O.}},
\bauthor{\bsnm{Duato}, \binits{J.}}:
\batitle{{A new proposal to deal with congestion in InfiniBand-based
  fat-trees}}.
\bjtitle{J. Parallel Distrib. Comput.}
\bvolume{74}(\bissue{1}),
\bfpage{1802}--\blpage{1819}
(\byear{2014}).
\doiurl{10.1016/j.jpdc.2013.09.002}
\end{barticle}
\endbibitem

\bibitem{MellanoxQuantum}
\begin{botherref}
\oauthor{\bsnm{Mellanox}}:
{NVIDIA MELLANOX QUANTUM -PRODUCT BRIEF}.
\url{{https://network.nvidia.com/sites/default/files/doc-2020/pb-quantum-hdr-switch-silicon.pdf}}
Accessed 2020-09-15
\end{botherref}
\endbibitem

\bibitem{MellanoxConfigureAR}
\begin{botherref}
\oauthor{\bsnm{Mellanox}}:
{How To Configure Adaptive Routing and SHIELD (New)}.
\url{https://support.mellanox.com/s/article/How-To-Configure-Adaptive-Routing-and-Self-Healing-Networking-New}
Accessed 2021-08-05
\end{botherref}
\endbibitem

\bibitem{DBLP:conf/ipps/Zahavi11}
\begin{bchapter}
\bauthor{\bsnm{Zahavi}, \binits{E.}}:
\bctitle{Fat-trees routing and node ordering providing contention free traffic
  for {MPI} global collectives}.
In: \bbtitle{25th {IEEE} International Symposium on Parallel and Distributed
  Processing, {IPDPS} 2011, Anchorage, Alaska, USA, 16-20 May 2011 - Workshop
  Proceedings},
pp. \bfpage{761}--\blpage{770}.
\bpublisher{{IEEE}}, \blocation{???}
(\byear{2011}).
\doiurl{10.1109/IPDPS.2011.219}.
\burl{http://dx.doi.org/10.1109/IPDPS.2011.219}
\end{bchapter}
\endbibitem

\bibitem{slimmedFT}
\begin{bchapter}
\bauthor{\bsnm{Rodr{\'{\i}}guez}, \binits{G.}},
\bauthor{\bsnm{Minkenberg}, \binits{C.}},
\bauthor{\bsnm{Beivide}, \binits{R.}},
\bauthor{\bsnm{Luijten}, \binits{R.P.}},
\bauthor{\bsnm{Labarta}, \binits{J.}},
\bauthor{\bsnm{Valero}, \binits{M.}}:
\bctitle{Oblivious routing schemes in extended generalized fat tree networks}.
In: \bbtitle{Proceedings of the 2009 {IEEE} International Conference on Cluster
  Computing, August 31 - September 4, 2009, New Orleans, Louisiana, {USA}},
pp. \bfpage{1}--\blpage{8}.
\bpublisher{{IEEE} Computer Society}, \blocation{???}
(\byear{2009}).
\doiurl{10.1109/CLUSTR.2009.5289145}.
\burl{https://doi.org/10.1109/CLUSTR.2009.5289145}
\end{bchapter}
\endbibitem

\bibitem{Zahavi10cpe}
\begin{barticle}
\bauthor{\bsnm{Zahavi}, \binits{E.}},
\bauthor{\bsnm{Johnson}, \binits{G.}},
\bauthor{\bsnm{Kerbyson}, \binits{D.J.}},
\bauthor{\bsnm{Lang}, \binits{M.}}:
\batitle{{Optimized InfiniBand fat-tree routing for shift all-to-all
  communication patterns}}.
\bjtitle{Concurrency and Computation: Practice and Experience}
\bvolume{22}(\bissue{2}),
\bfpage{217}--\blpage{231}
(\byear{2010}).
\doiurl{10.1002/cpe.1527}
\end{barticle}
\endbibitem

\bibitem{OmnipathFM}
\begin{botherref}
Intel® omni-path fabric suite fabric manager.
(2015).
\url{https://www.intel.com/content/dam/support/us/en/documents/network/omni-adptr/sb/Intel_OP_FabricSuite_Fabric_Manager_UG_H76468_v1_0.pdf}
\end{botherref}
\endbibitem

\bibitem{BXIV2}
\begin{botherref}
\oauthor{\bsnm{Vignéras}, \binits{P.}},
\oauthor{\bsnm{Quintin}, \binits{J.-N.}}:
The bxi routing architecture for exascale supercomputer.
The Journal of Supercomputing
\textbf{72}
(2016).
\doiurl{10.1007/s11227-016-1755-2}
\end{botherref}
\endbibitem

\bibitem{slingshot}
\begin{bchapter}
\bauthor{\bsnm{De~Sensi}, \binits{D.}},
\bauthor{\bsnm{Di~Girolamo}, \binits{S.}},
\bauthor{\bsnm{McMahon}, \binits{K.}},
\bauthor{\bsnm{Roweth}, \binits{D.}},
\bauthor{\bsnm{Hoefler}, \binits{T.}}:
\bctitle{An in-depth analysis of the slingshot interconnect},
pp. \bfpage{1}--\blpage{14}
(\byear{2020}).
\doiurl{10.1109/SC41405.2020.00039}
\end{bchapter}
\endbibitem

\bibitem{IBAspec2015}
\begin{botherref}
\oauthor{\bsnm{{InfiniBand Trade Association.}}}:
{InfiniBandTM} {Architecture} {Specification} {Volume} 1 - {Release} 1.3
(2015)
\end{botherref}
\endbibitem

\bibitem{patenteARN}
\begin{botherref}
\oauthor{\bsnm{Haramaty}, \binits{Z.}},
\oauthor{\bsnm{Zahavi}, \binits{E.}},
\oauthor{\bsnm{Gabbay}, \binits{F.}},
\oauthor{\bsnm{Crupnicoff}, \binits{D.}},
\oauthor{\bsnm{Marelli}, \binits{A.}},
\oauthor{\bsnm{Bloch}, \binits{G.}}:
Adaptive Routing Using Inter-switch Notifications.
US20140211631A1,
Apr 2015
\end{botherref}
\endbibitem

\bibitem{GranZRSSL11}
\begin{bchapter}
\bauthor{\bsnm{Gran}, \binits{E.G.}},
\bauthor{\bsnm{Zahavi}, \binits{E.}},
\bauthor{\bsnm{Reinemo}, \binits{S.}},
\bauthor{\bsnm{Skeie}, \binits{T.}},
\bauthor{\bsnm{Shainer}, \binits{G.}},
\bauthor{\bsnm{Lysne}, \binits{O.}}:
\bctitle{On the relation between congestion control, switch arbitration and
  fairness}.
In: \bbtitle{11th {IEEE/ACM} International Symposium on Cluster, Cloud and Grid
  Computing, CCGrid 2011, Newport Beach, CA, USA, May 23-26, 2011},
pp. \bfpage{342}--\blpage{351}.
\bpublisher{{IEEE} Computer Society}, \blocation{???}
(\byear{2011}).
\doiurl{10.1109/CCGrid.2011.67}.
\burl{https://doi.org/10.1109/CCGrid.2011.67}
\end{bchapter}
\endbibitem

\bibitem{ESCUDEROSAHUQUILLO201835}
\begin{barticle}
\bauthor{\bsnm{Escudero-Sahuquillo}, \binits{J.}},
\bauthor{\bsnm{Garcia}, \binits{P.J.}},
\bauthor{\bsnm{Quiles}, \binits{F.J.}},
\bauthor{\bsnm{Maglione-Mathey}, \binits{G.}},
\bauthor{\bsnm{Duato}, \binits{J.}}:
\batitle{Feasible enhancements to congestion control in infiniband-based
  networks}.
\bjtitle{Journal of Parallel and Distributed Computing}
\bvolume{112},
\bfpage{35}--\blpage{52}
(\byear{2018}).
\doiurl{10.1016/j.jpdc.2017.09.008}
\end{barticle}
\endbibitem

\bibitem{Yebenes13PDP}
\begin{bchapter}
\bauthor{\bsnm{Yebenes}, \binits{P.}},
\bauthor{\bsnm{Escudero{-}Sahuquillo}, \binits{J.}},
\bauthor{\bsnm{Garc{\'{\i}}a}, \binits{P.J.}},
\bauthor{\bsnm{Quiles}, \binits{F.J.}}:
\bctitle{Towards modeling interconnection networks of exascale systems with
  omnet++}.
In: \bbtitle{21st Euromicro International Conference on Parallel, Distributed,
  and Network-Based Processing, {PDP} 2013, Belfast, United Kingdom, February
  27 - March 1, 2013},
pp. \bfpage{203}--\blpage{207}
(\byear{2013}).
\doiurl{10.1109/PDP.2013.36}
\end{bchapter}
\endbibitem

\bibitem{OMNeTweb}
\begin{botherref}
\oauthor{\bsnm{{OpenSim Ltd}}}:
{OMNeT++ Discrete Event Simulator}
\end{botherref}
\endbibitem

\bibitem{GranCCgrid}
\begin{bchapter}
\bauthor{\bsnm{Gran}, \binits{E.G.}},
\bauthor{\bsnm{Zahavi}, \binits{E.}},
\bauthor{\bsnm{Reinemo}, \binits{S.-A.}},
\bauthor{\bsnm{Skeie}, \binits{T.}},
\bauthor{\bsnm{Shainer}, \binits{G.}},
\bauthor{\bsnm{Lysne}, \binits{O.}}:
\bctitle{On the relation between congestion control, switch arbitration and
  fairness}.
In: \bbtitle{2011 11th IEEE/ACM International Symposium on Cluster, Cloud and
  Grid Computing},
pp. \bfpage{342}--\blpage{351}
(\byear{2011}).
\doiurl{10.1109/CCGrid.2011.67}
\end{bchapter}
\endbibitem

\bibitem{Andujar16JSC}
\begin{barticle}
\bauthor{\bsnm{Andujar}, \binits{F.J.}},
\bauthor{\bsnm{Villar}, \binits{J.A.}},
\bauthor{\bsnm{Alfaro}, \binits{F.J.}},
\bauthor{\bsnm{S{\'{a}}nchez}, \binits{J.L.}},
\bauthor{\bsnm{Escudero{-}Sahuquillo}, \binits{J.}}:
\batitle{An open-source family of tools to reproduce mpi-based workloads in
  interconnection network simulators}.
\bjtitle{The Journal of Supercomputing}
\bvolume{72}(\bissue{12}),
\bfpage{4601}--\blpage{4628}
(\byear{2016})
\end{barticle}
\endbibitem

\bibitem{PTRANS}
\begin{botherref}
{The HPCC Benchmark}.
http://icl.cs.utk.edu/hpcc/.
\url{http://icl.cs.utk.edu/hpcc/}
Accessed 2016-12-19
\end{botherref}
\endbibitem

\bibitem{inception}
\begin{bchapter}
\bauthor{\bsnm{Szegedy}, \binits{C.}},
\bauthor{\bsnm{Vanhoucke}, \binits{V.}},
\bauthor{\bsnm{Ioffe}, \binits{S.}},
\bauthor{\bsnm{Shlens}, \binits{J.}},
\bauthor{\bsnm{Wojna}, \binits{Z.}}:
\bctitle{Rethinking the inception architecture for computer vision}.
In: \bbtitle{2016 {IEEE} Conference on Computer Vision and Pattern Recognition,
  {CVPR} 2016, Las Vegas, NV, USA, June 27-30, 2016},
pp. \bfpage{2818}--\blpage{2826}.
\bpublisher{{IEEE} Computer Society}, \blocation{???}
(\byear{2016}).
\doiurl{10.1109/CVPR.2016.308}.
\burl{https://doi.org/10.1109/CVPR.2016.308}
\end{bchapter}
\endbibitem

\end{thebibliography}


\end{document}